\documentclass[lettersize,journal]{IEEEtran}
\usepackage{amsmath, amssymb, amsfonts}

\usepackage{algorithm}
\usepackage{algorithmic}

\usepackage{array}
\usepackage{tabularx}
\usepackage{booktabs}
\usepackage{colortbl}

\usepackage{graphicx}
\usepackage{caption}

\usepackage[dvipsnames]{xcolor}

\usepackage{cite}
\usepackage{url}
\usepackage{textcomp}
\usepackage{verbatim}
\usepackage{flushend}
\usepackage{bookmark}

\usepackage{pgfplots}
\usepackage{pgfplotstable}
\pgfplotsset{compat=1.18}
\usepgfplotslibrary{groupplots}

\usepackage{tikz}
\usetikzlibrary{positioning, fit, arrows.meta, calc, plotmarks}

\newcommand{\csvPath}{results/diverse}

\pgfplotstableread[col sep=comma]{\csvPath/nmse_steps_snr_0dB.csv}\diverseZero
\pgfplotstableread[col sep=comma]{\csvPath/nmse_steps_snr_10dB.csv}\diverseTen
\pgfplotstableread[col sep=comma]{\csvPath/nmse_steps_snr_20dB.csv}\diverseTwenty
\pgfplotstableread[col sep=comma]{\csvPath/se_steps_snr_0dB.csv}\diverseZeroSe
\pgfplotstableread[col sep=comma]{\csvPath/se_steps_snr_10dB.csv}\diverseTenSe
\pgfplotstableread[col sep=comma]{\csvPath/se_steps_snr_20dB.csv}\diverseTwentySe
\pgfplotstableread[col sep=comma]{\csvPath/nmse_summary_first_last_avg.csv}\diversesummary
\pgfplotstableread[col sep=comma]{\csvPath/nmse_steps_by_Tin.csv}\Context
\pgfplotstableread[col sep=comma]{\csvPath/mmse_nmse_vs_snr_perSample_minmax.csv}\mmseTable
\pgfplotstableread[col sep=comma]{\csvPath/3GHz_test.csv}\threeGHz

\newcommand{\pathThirty}{results/30}
\newcommand{\pathSixty}{results/60}
\newcommand{\pathOneTwenty}{results/120}

\def\snrlabel{10dB}

\pgfplotstableread[col sep=comma]{\pathThirty/nmse_steps_snr_\snrlabel.csv}\dataThirty
\pgfplotstableread[col sep=comma]{\pathSixty/nmse_steps_snr_\snrlabel.csv}\dataSixty
\pgfplotstableread[col sep=comma]{\pathOneTwenty/nmse_steps_snr_\snrlabel.csv}\dataOneTwenty

\def\modellist{Diffusion, 30vs10, LinFusion, LinFormer,LSTM,GRU}
\def\modellistSE{Diu, 30vs10, LinFusion, LinFormer,Conv-LSTM,GRU}

\pgfplotscreateplotcyclelist{mymarkers}{
  {color=MidnightBlue, mark=square*,           mark size=1.pt, line width=0.3pt},
  {color=BrickRed,     mark=halfcircle*,     mark size=1.5pt, line width=0.3pt},
  {color=ForestGreen,  mark=diamond*,   mark size=1.5pt, line width=0.3pt},
  {color=Purple,       mark=star,    mark size=1.5pt, line width=0.3pt, dashed},
  {color=BurntOrange,  mark=triangle*,           mark size=1.5pt, line width=0.3pt, dashed},
  {color=Black,        mark=oplus*, mark size=1.pt, line width=0.3pt, dashed},
  {color= Magenta,          mark=halfcircle*,semithick, mark size=1.5pt, line width=0.3pt},
  {color=Cyan,         mark=pentagon*,semithick, mark size=1.5pt, line width=0.3pt},
  {color= OliveGreen,       mark=oplus*,   semithick, mark size=1.5pt, line width=0.3pt},
  {color=Gray,            mark=diamond*,     semithick, mark size=1.5pt, line width=0.3pt},
  {color=Cerulean,        mark=halfcircle*, semithick, mark size=1.5pt, line width=0.3pt},
}

\pgfplotscreateplotcyclelist{mymarkers2}{
  {color=MidnightBlue, mark=oplus*,           mark size=1.pt, line width=0.3pt},
  {color=BrickRed,     mark=square*,     mark size=1.pt, line width=0.3pt},
  {color=ForestGreen,  mark=triangle*,   mark size=1.5pt, line width=0.3pt},
  {color=Purple,       mark=star,    mark size=1.5pt, line width=0.3pt},
  {color=BurntOrange,  mark=|,           mark size=1.5pt, line width=0.3pt},
  {color=Black,        mark=pentagon*, mark size=1.0pt, line width=0.3pt},
  {color= Magenta,          mark=halfcircle*,semithick, mark size=1.5pt, line width=0.3pt},
  {color=Cyan,         mark=pentagon*,semithick, mark size=1.5pt, line width=0.3pt},
  {color= OliveGreen,       mark=oplus*,   semithick, mark size=1.5pt, line width=0.3pt},
  {color=Gray,            mark=diamond*,     semithick, mark size=1.5pt, line width=0.3pt},
  {color=Cerulean,        mark=halfcircle*, semithick, mark size=1.5pt, line width=0.3pt},
}

\begin{document}

\title{CSI Prediction Using Diffusion Models}

\author{Mehdi~Sattari,~\IEEEmembership{Graduate~Student~Member,~IEEE,}
        Javad~Aliakbari,~\IEEEmembership{Graduate~Student~Member,~IEEE,}
        Alexandre~Graell~i~Amat,~\IEEEmembership{Senior~Member,~IEEE,}
        and
        Tommy~Svensson,~\IEEEmembership{Senior~Member,~IEEE}
        ~
\thanks{This work was supported in part by the SEMANTIC project, funded by the EU’s Horizon 2020 research and innovation programme under the Marie Skłodowska-Curie grant agreement No.~861165, and in part by the Hexa-X-II project, which has received funding from the Smart Networks and Services Joint Undertaking (SNS JU) under the European Union’s Horizon Europe research and innovation programme under grant agreement No.~101095759. This work was also partially supported by the Swedish Research Council (VR) under grants 2020-03687 and 2023-05065, by the Wallenberg AI, Autonomous Systems and Software Program (WASP) funded by the Knut and Alice Wallenberg Foundation, and by the Swedish Innovation Agency (Vinnova) under grant 2022-03063. The computations were enabled by resources provided by the National Academic Infrastructure for Supercomputing in Sweden (NAISS), partially funded by the Swedish Research Council through grant agreement No.~2022-06725.
}
}

\maketitle

\begin{abstract}

Acquiring accurate channel state information (CSI) is critical for reliable and efficient wireless communication, but challenges such as high pilot overhead and channel aging hinder timely and accurate CSI acquisition. CSI prediction, which forecasts future CSI from historical observations, offers a promising solution. Recent deep learning approaches, including recurrent neural networks and Transformers, have achieved notable success but typically learn deterministic mappings, limiting their ability to capture the stochastic and multimodal nature of wireless channels. In this paper, we introduce a novel probabilistic framework for CSI prediction based on diffusion models, offering a flexible design that supports integration of diverse prediction schemes. We decompose the CSI prediction task into two components: a temporal encoder, which extracts channel dynamics, and a diffusion-based generator, which produces future CSI samples. We investigate two inference schemes—autoregressive and sequence-to-sequence—and explore multiple diffusion backbones, including U-Net and Transformer-based architectures. Furthermore, we examine a diffusion-based approach without an explicit temporal encoder and utilize the DDIM scheduling to reduce model complexity. Extensive simulations demonstrate that our diffusion-based models significantly outperform state-of-the-art baselines.

\end{abstract}


\section{Introduction}\label{sec1}
\IEEEPARstart{M}{}ultiple-input multiple-output (MIMO) technology plays a pivotal role in advanced wireless communications. To fully exploit the benefits of MIMO systems, accurate acquisition of channel state information (CSI) is essential for designing efficient precoding and combining algorithms. However, CSI acquisition is challenged by high overhead, channel aging, and significant computational complexity. To address these issues, a variety of algorithms have been proposed to obtain CSI with reduced overhead and complexity \cite{Biglieri2007,1193803,6736746,6847111,4389761}.

CSI prediction refers to the task of forecasting future CSI based on historical observations. Various classical approaches have been proposed, including linear extrapolation \cite{9127447}, Kalman filtering \cite{9210016}, sum-of-sinusoids models \cite{1578065}, Gaussian processes \cite{7275189}, predictor antenna \cite{9502657}, and autoregressive (AR) models \cite{1512123}. However, the underlying statistics of wireless channels are inherently complex and difficult to model accurately, making reliable CSI prediction particularly challenging. As a result, traditional methods often fail to deliver satisfactory performance, especially under highly dynamic channel conditions.

Deep neural networks (DNNs) have been increasingly adopted in wireless communication tasks \cite{9410430}, \cite{9037126}.
Motivated by their significant success in time-series forecasting problems \cite{FISCHER2018654, 8039509, ijcai2018p505, meijer2024risediffusionmodelstimeseries, rasul2021autoregressivedenoisingdiffusionmodels}, the use of DNNs for CSI prediction has attracted growing research interest \cite{9427230, 10422880, 9921297, 8746352, 9044427, 10978424, kim2025machinelearningfuturewireless, 9814839, 9832933, 10965849, 10706231, akrout2024nextslotofdmcsipredictionmultihead, 10363354, zhao2025csibert2bertinspiredframeworkefficient, 10457056}.
Various neural network (NN) architectures have been explored to capture the temporal dynamics of CSI, including recurrent neural networks (RNNs), such as long short-term memory (LSTM) and gated recurrent units (GRUs), as well as Transformer-based models. RNN-based architectures have been successfully employed for CSI prediction in \cite{8746352, 9044427, 9814839}, while Transformer-based approaches have been investigated in \cite{9832933, 10965849}, showing superior performance in modeling long-range temporal dependencies.

Accurately predicting CSI is a fundamentally challenging problem due to the highly stochastic nature of wireless channels and their complex underlying distributions. Future channel states are influenced by a wide range of factors, including multipath fading, user mobility, and environmental dynamics, which together lead to significant uncertainty in CSI evolution. Deep learning models such as RNNs and Transformers have shown promising results for time-series forecasting, but they primarily learn a deterministic mapping from historical CSI to future CSI. This approach often fails to capture the inherent randomness and multimodality of wireless channel evolution.
In contrast, recent advances in generative artificial intelligence (AI) have enabled the development of models that learn the full underlying probability distribution of data. Among these, diffusion models \cite{NEURIPS2020_4c5bcfec} have achieved state-of-the-art performance in high-dimensional data generation tasks across various modalities. By modeling CSI prediction as a probabilistic generative process, diffusion models can naturally account for uncertainty in channel evolution while learning the joint spatiotemporal structure of CSI sequences.

Motivated by the success of recent advances of diffusion models in modeling complex data distributions and the limitations of RNN- and Transformer-based approaches in CSI prediction, this paper explores the application of diffusion models to the CSI prediction problem. The main contributions of this work are summarized as follows:
\begin{itemize}
\item We propose a general generative framework for CSI prediction. The framework provides a flexible and modular design that enables seamless integration of diverse CSI prediction models. We formulate the CSI prediction task as two components: temporal encoding and generative sampling. The temporal encoder extracts latent representations of the channel dynamics, while the diffusion generator produces the next CSI frame conditioned on these latent features.

\item Within the proposed framework, we explore two inference schemes: AR and sequence-to-sequence (seq2seq). The AR strategy recursively predicts future CSI frames, enabling a single trained model to flexibly adapt to arbitrary context lengths and prediction horizons. In contrast, the seq2seq scheme enables parallel prediction of multiple future CSI frames in a single pass, albeit with reduced flexibility in handling variable context lengths and prediction horizons.
    
\item To reduce complexity for real-time applications, we investigate a simplified design in which the diffusion model directly learns temporal and spatial dependencies without an explicit temporal encoder. Furthermore, to enhance sampling efficiency and stability, we adopt the denoising diffusion implicit model (DDIM) \cite{song2022denoisingdiffusionimplicitmodels} scheduling technique. Our experiments show that as few as three sampling steps are sufficient to achieve competitive performance.

\item We validate the generality of the framework by integrating multiple generator backbones, including U-Net \cite{ronneberger2015unetconvolutionalnetworksbiomedical}, 3D U-Net \cite{Cicek2016_3DUNet}, and diffusion Transformer (DiT) \cite{peebles2023scalablediffusionmodelstransformers}, in combination with temporal encoders such as ConvLSTM \cite{shi2015convolutionallstmnetworkmachine} and LinFormer \cite{10965849}. Extensive simulation results demonstrate that the proposed diffusion-based framework significantly outperforms state-of-the-art baselines and exhibits strong generalization to unseen wireless environments.
\end{itemize}

The rest of this paper is organized as follows: Section \ref{sec2} introduces the channel model and CSI prediction problem, as well as a preliminary for diffusion models. In Section \ref{sec3}, we present our diffusion-based CSI prediction framework and training and inference procedures. Section \ref{sec4} provides numerical simulations and compares different CSI prediction schemes in terms of the normalized mean squared error (NMSE). Finally, Section \ref{sec5} concludes the paper.

\section{Preliminaries and Problem Formulation}\label{sec2}
\subsection{Channel Model}\label{CDL_model}
We focus on the clustered delay line (CDL) channel model based on the 3GPP specification \cite{3gpp.38.901}. The CDL channel model is a spatial channel model for wireless channel simulation. This model comprises different scenarios, denoted as \texttt{CDL-A} to \texttt{CDL-E}, where each scenario models different channel statistics with different delay profiles, angular spreads, etc. For example, \texttt{CDL-B} generates an urban macrocell scenario with a wider angular spread.
The time-varying MIMO channel impulse response in the CDL model is given by
\begin{equation}
\mathbf{H}(t) = \sum_{l=1}^{L} \sum_{r=1}^{R} \alpha_{l,r}(t) \,
\mathbf{a}_\text{Rx}\left(\phi_{l,r}^{\text{Rx}}\right) \,
\mathbf{a}_\text{Tx}^H\left(\phi_{l,r}^{\text{Tx}}\right) \,
e^{-j 2\pi f \tau_l},
\end{equation}
or its discrete counterpart,
\begin{equation}
\mathbf{H}_n = \sum_{l=1}^{L} \sum_{r=1}^{R} \alpha_{l,r}[n] \,
\mathbf{a}_\text{Rx}\left(\phi_{l,r}^{\text{Rx}}\right) \,
\mathbf{a}_\text{Tx}^H\left(\phi_{l,r}^{\text{Tx}}\right) \,
e^{-j 2\pi \frac{k f_s}{N_\mathrm{c}} \tau_l},
\end{equation}
where,
$L$ is the number of clusters,
$R$ the number of rays per cluster,
$\alpha_{l,r}(t)$ time-varying complex path gain for ray $r$ in cluster $l$,
$\mathbf{a}_\text{Rx}(\cdot), \mathbf{a}_\text{Tx}(\cdot)$ denote receive/transmit array response vectors,
$(\phi^\text{Rx}, \phi^\text{Tx})$ represent angles of arrival/departure,
$f_s$ the sampling frequency,
$N_c$ the number of subcarriers,
and $\tau_l$ is the delay of the $l$-th cluster. 
The time-varying complex path gain is modeled as
\begin{equation}
\alpha_{l,r}[n] = \sqrt{P_{l,r}} \, e^{j (2\pi f_{l,r} n + \phi_{l,r})},
\end{equation}
where
$P_{l,r}$ is the power of the $r$-th ray in cluster $l$,
$f_{l,r}$ the Doppler shift, and
$\phi_{l,r}$ the random initial phase.
For a uniform linear array (ULA), the array response vector is
\begin{equation}
\mathbf{a}(\phi) = \frac{1}{\sqrt{N_\mathrm{t}}} \left[1, \, e^{j 2\pi \frac{d}{\lambda} \sin(\phi)}, \, \dots, \, e^{j 2\pi \frac{d}{\lambda}(N_\mathrm{t}-1) \sin(\phi)} \right]^T,
\end{equation}
where,
$N_\mathrm{t}$ denotes the number of antenna elements and
$d$ the antenna spacing.

We consider a MIMO system in which a base station (BS) serves multiple single-antenna users. The BS is equipped with a ULA of \( N_\mathrm{t} \) antennas. Orthogonal frequency division multiplexing (OFDM) with \( N_\mathrm{c} \) subcarriers is employed for downlink transmission.
For each subcarrier $m \in \{1,\ldots,N_\mathrm{c}\}$ and time index $n$, 
$\mathbf{h}_{m,n} \in \mathbb{C}^{N_\mathrm{t}}$ denotes the channel vector from the BS  with $N_\mathrm{t}$ antennas to the user. By stacking all subcarriers, the downlink CSI 
matrix in the spatial--frequency domain at time $n$ is given by
\begin{equation}
    \mathbf{H}_n = \big[\mathbf{h}_{1,n} \; \mathbf{h}_{2,n} \; \cdots \; \mathbf{h}_{N_\mathrm{c},n}\big] 
    \in \mathbb{C}^{N_\mathrm{t} \times N_\mathrm{c}}.
\end{equation}

\subsection{Diffusion Models}
\paragraph{Denoising diffusion probabilistic models (DDPMs)}Diffusion models are a class of generative models that learn to approximate complex data distributions by progressively corrupting the data with noise and then reversing this process. The foundational work on DDPM was introduced in \cite{NEURIPS2020_4c5bcfec}. In DDPMs, the forward process gradually adds Gaussian noise to the data, while the reverse process aims to reconstruct the original data using a NN. This is modeled as a Markov chain with a predefined, time-dependent noise schedule.

The forward process defines a Markov chain that incrementally perturbs the data over \( T \) time steps. Starting from a data sample \( \mathbf{H}^{0} \sim p(\mathbf{H}^{0}) \), the forward diffusion is expressed as
\begin{equation}
    q(\mathbf{H}^{1:T} \mid \mathbf{H}^{0}) = \prod_{t=1}^{T} q(\mathbf{H}^{t} \mid \mathbf{H}^{t-1}),
\end{equation}
where each conditional distribution is Gaussian
\begin{equation}
    q(\mathbf{H}^{t} \mid \mathbf{H}^{t-1}) = \mathcal{N} \left( \mathbf{H}^t; \sqrt{1 - \beta_t} \, \mathbf{H}^{t-1}, \beta_t \, \mathbf{I} \right),
\end{equation}
with \( \beta_t \in (0, 1) \) denoting the noise variance at step \( t \), and \( \mathbf{H}^t \) representing the noisy latent at time \( t \).

The marginal distribution at an arbitrary time step \( t \), conditioned directly on the original sample \( \mathbf{H}^0 \), is
\begin{equation}
    q(\mathbf{H}^t \mid \mathbf{H}^0) = \mathcal{N} \left( \mathbf{H}^t; \sqrt{\bar{\alpha}_t} \, \mathbf{H}^0, (1 - \bar{\alpha}_t) \, \mathbf{I} \right),
\end{equation}
which leads to the reparameterization
\begin{equation}\label{forward_diffusion_transition}
    \mathbf{H}^t = \sqrt{\bar{\alpha}_t} \, \mathbf{H}^0 + \sqrt{1 - \bar{\alpha}_t} \, \boldsymbol{\epsilon}, \quad \boldsymbol{\epsilon} \sim \mathcal{N}(\mathbf{0}, \mathbf{I}),
\end{equation}
where \( \boldsymbol{\epsilon} \) denotes standard Gaussian noise, and
\begin{equation}
    \bar{\alpha}_t = \prod_{i=1}^t \alpha_i, \quad \text{with} \quad \alpha_t \triangleq 1 - \beta_t.
\end{equation}

The noise schedule is crafted such that \( \bar{\alpha}_T \approx 0 \), ensuring that the final latent \( \mathbf{H}^T \) is nearly standard Gaussian.

The reverse process is a learned Markov chain that denoises \( \mathbf{H}^T \sim \mathcal{N}(\mathbf{0}, \mathbf{I}) \) back to the data domain, and each reverse transition is approximated as
\begin{equation}
    p_{\boldsymbol{\theta}}(\mathbf{H}^{t-1} \mid \mathbf{H}^t) = \mathcal{N} \left( \mathbf{H}^{t-1}; \boldsymbol{\mu}_{\boldsymbol{\theta}}(\mathbf{H}^t, t), \boldsymbol{\Sigma}_{\boldsymbol{\theta}}(t) \right),
\end{equation}
with the prior given by
\begin{equation}
    p(\mathbf{H}^T) = \mathcal{N}(\mathbf{H}^T; \mathbf{0}, \mathbf{I}).
\end{equation}

In DDPM, the mean \( \boldsymbol{\mu}_{\boldsymbol{\theta}}(\mathbf{H}^t, t) \) is expressed using a predicted noise function \( \boldsymbol{\epsilon}_{\boldsymbol{\theta}} \)
\begin{equation}
    \boldsymbol{\mu}_{\boldsymbol{\theta}}(\mathbf{H}^t, t) = \frac{1}{\sqrt{\alpha_t}} \left( \mathbf{H}^t - \frac{1 - \alpha_t}{\sqrt{1 - \bar{\alpha}_t}} \, \boldsymbol{\epsilon}_{\boldsymbol{\theta}}(\mathbf{H}^t, t) \right),
\end{equation}
where \( \boldsymbol{\epsilon}_{\boldsymbol{\theta}}(\mathbf{H}^t, t) \) is an NN trained to predict the forward noise at time \( t \).
The variance \( \boldsymbol{\Sigma}_{\boldsymbol{\theta}}(t) \) is usually fixed and time-dependent
\begin{equation}
    \boldsymbol{\Sigma}_{\boldsymbol{\theta}}(t) = \tilde{\beta}_t \, \mathbf{I}, \quad \text{with} \quad \tilde{\beta}_t = \frac{1 - \bar{\alpha}_{t-1}}{1 - \bar{\alpha}_t} \, \beta_t.
\end{equation}

The model is trained by minimizing a simplified variational lower bound, which reduces to a weighted mean squared error (MSE) between the true noise \( \boldsymbol{\epsilon} \) and the predicted noise \( \boldsymbol{\epsilon}_{\boldsymbol{\theta}} \)
\begin{equation}
    \mathcal{L}_{\text{noise}} = \mathbb{E}_{\mathbf{H}^0, t, \boldsymbol{\epsilon}} \left[ \left\| \boldsymbol{\epsilon} - \boldsymbol{\epsilon}_{\boldsymbol{\theta}}(\mathbf{H}^t, t) \right\|^2 \right].
\end{equation}

An alternative yet equivalent training objective is to have the model directly predict the clean data \( \mathbf{H}^0 \) from the noisy observation \( \mathbf{H}^t \). In this case, the training loss becomes the MSE between the true clean sample and the predicted sample \( \mathbf{H}^{0}_{\boldsymbol{\theta}} \)
\begin{equation}\label{loss_diff_sample}
    \mathcal{L}_{\text{data}} = \mathbb{E}_{\mathbf{H}^0, t, \boldsymbol{\epsilon}} \left[ \left\| \mathbf{H}^0 - \mathbf{H}^0_{\boldsymbol{\theta}}(\mathbf{H}^t, t) \right\|^2 \right].
\end{equation}

This formulation is particularly useful in inverse problems, such as denoising or super-resolution, where the objective is to recover a clean signal rather than generate diverse samples. Both loss functions are mathematically connected through the forward diffusion equation, and the predicted clean sample can be converted to noise (or vice versa) using closed-form expressions.

\paragraph{DDIM} The DDPM framework involves a stochastic reverse process and often requires a large number of diffusion steps (e.g., \(T = 1000\)) to achieve high-quality generation. To reduce the number of steps without significant degradation in quality, Song~\emph{et al.}~\cite{song2022denoisingdiffusionimplicitmodels} proposed DDIM, which reinterprets the reverse process as a non-Markovian mapping that can be made fully deterministic.
Starting from the DDPM estimate of the clean sample \(\hat{\mathbf{H}}^0\) at time step \(t\)
\begin{equation}
    \hat{\mathbf{H}}^0 = \frac{1}{\sqrt{\bar{\alpha}_t}} \left( \mathbf{H}^t - \sqrt{1 - \bar{\alpha}_t} \, \boldsymbol{\epsilon}_{\boldsymbol{\theta}}(\mathbf{H}^t, t) \right),
\end{equation}
and the DDIM update is given by
\begin{equation}
    \mathbf{H}^{t-1} = \sqrt{\bar{\alpha}_{t-1}} \, \hat{\mathbf{H}}^0 
    + \sqrt{1 - \bar{\alpha}_{t-1} - \sigma_t^2} \, \boldsymbol{\epsilon}_{\boldsymbol{\theta}}(\mathbf{H}^t, t) 
    + \sigma_t \, \boldsymbol{\epsilon}, 
\end{equation}
where \(\boldsymbol{\epsilon} \sim \mathcal{N}(0, \mathbf{I})\) and 
\begin{equation}
\sigma_t = \zeta \cdot \sqrt{\frac{1 - \bar{\alpha}_{t-1}}{1 - \bar{\alpha}_t}} 
\cdot \sqrt{1 - \frac{\bar{\alpha}_t}{\bar{\alpha}_{t-1}}},
\end{equation}
where \(\zeta \in [0,1]\) controls the level of stochasticity, \(\zeta = 0\) yields a fully deterministic process, while \(\zeta = 1\) recovers the stochastic DDPM update.
This formulation eliminates the need to sample fresh Gaussian noise at each step in the deterministic setting (\(\zeta = 0\)), enabling fast sampling and allowing for arbitrary step schedules. By tuning \(\zeta\), DDIM provides a trade-off between sample diversity and generation speed.

\section{CSI Prediction}\label{sec3}
\subsection{{Problem Formulation}}
Time-varying wireless channels can be naturally modeled as time-series data, where CSI prediction entails capturing both the spatial distributions and temporal dynamics of the channel. Under the MSE criterion, the optimal predictor corresponds to the minimum mean squared error (MMSE) estimate, which can be formulated as
\begin{equation}
    f^\ast (\mathbf{H}_{\mathrm{p}})
    = \arg\min_{f} \ 
    \mathbb{E}\!\left[\lVert \mathbf{H}_{\mathrm{f}} - f(\mathbf{H}_{\mathrm{p}}) \rVert^2 \right],
\end{equation}
where 
\(\mathbf{H}_{\mathrm{f}} \triangleq \left\{ \mathbf{H}_{n+1}, \mathbf{H}_{n+2}, \dots, \mathbf{H}_{n+N_{\mathrm{f}}} \right\}\) 
denotes the set of future CSI over the next \(N_{\mathrm{f}}\) time steps, and 
\(\mathbf{H}_{\mathrm{p}} \triangleq \left\{ \mathbf{H}_{n-N_{\mathrm{p}}+1}, \mathbf{H}_{n-N_{\mathrm{p}}+2}, \dots, \mathbf{H}_{n} \right\}\) 
represents the historical CSI observations over the previous \(N_{\mathrm{p}}\) time steps. The optimal predictor \(f^\ast (\mathbf{H}_{\mathrm{p}})\) is the conditional mean estimator (CME)~\cite{fesl2025asymptoticmeansquareerror}, given by
\begin{equation}\label{eq: CME}
\begin{aligned}
    f^\ast (\mathbf{H}_{\mathrm{p}}) 
    &= \mathbb{E}[\mathbf{H}_{\mathrm{f}} \mid \mathbf{H}_{\mathrm{p}}] \\[4pt]
    &= \int \mathbf{H}_{\mathrm{f}} \, p(\mathbf{H}_{\mathrm{f}} \mid \mathbf{H}_{\mathrm{p}}) \, d\mathbf{H}_{\mathrm{f}}. \\[4pt]
\end{aligned}
\end{equation}

Deriving the optimal MMSE predictor requires knowledge of both the conditional distribution of the future CSI given the past (capturing the channel dynamics) and the prior distribution of the wireless channel (capturing the spatial statistics). Deriving this predictor in closed form is intractable due to high-dimensional integrals, nonlinear time-varying dynamics, and the lack of a known prior for practical channels.

A practical approach to approximate the optimal predictor is to employ discriminative DNNs, such as RNNs or Transformers. A discriminative DNN with learnable parameters \(\boldsymbol{\theta}\) aims to learn the nonlinear mapping from historical CSI observations, \(\mathbf{H}_{\mathrm{p}}\), to the future CSI, \(\mathbf{H}_{\mathrm{f}}\). With sufficient data and model capacity, a DNN trained using the MSE loss provides a data-driven approximation of the MMSE predictor, i.e.,
\begin{equation}
    f_{\boldsymbol{\theta}}(\mathbf{H}_{\mathrm{p}}) \approx f^\ast (\mathbf{H}_{\mathrm{p}}).
\end{equation}

However, such approximations face several limitations. First, jointly capturing the temporal dynamics of the wireless channel together with its spatial distribution is highly challenging. Second, discriminative models provide a deterministic mapping that fails to capture the uncertainty of channel evolution. Third, learning such complex dynamics from finite training data makes these models prone to overfitting, thereby limiting their robustness in practical deployments.

Furthermore, realistic wireless channels evolve in highly dynamic environments due to user mobility, multipath propagation, and environmental changes. As a result, the conditional distribution of future CSI given past observations can be highly uncertain and potentially multi-modal. Simple probabilistic models with restrictive distributional assumptions are therefore often insufficient to capture the full variability of CSI evolution.

Generative models provide a natural framework for learning such data distributions. In particular, diffusion models have recently demonstrated strong performance in modeling high-dimensional data without imposing restrictive parametric assumptions. This makes them a promising approach for capturing the stochastic evolution of wireless channels and enabling more robust CSI prediction. Moreover, diffusion models can be naturally extended to conditional settings by incorporating side information, such as past CSI observations, into the denoising process, enabling flexible and effective modeling of the conditional distribution of future channels.

\subsection{Diffusion CSI Prediction}

Inspired by the formulation of the optimal CME in~\eqref{eq: CME}, we decompose the prediction task into two components: a temporal encoder and a generator. The temporal encoder extracts latent temporal representations from the time-varying channel, while the generator learns the underlying channel distribution. To capture the probabilistic nature of CSI prediction, we employ diffusion models as the generator. In contrast to discriminative DNNs that produce deterministic point estimates, diffusion models yield probabilistic predictions, thereby accounting for the inherent uncertainty of wireless channels. Furthermore, since diffusion models are trained across the entire signal-to-noise ratio (SNR) range, they are expected to generalize more robustly to diverse channel conditions compared to task-specific discriminative models.

\begin{algorithm}[t]
\caption{Training}
\label{alg:training_procedure}
\textbf{Input:} 
Dataset $\mathcal{D} = \{(\mathbf{X}, \mathbf{Y})\}$; 
model parameters $\boldsymbol{\theta} = [\boldsymbol{\theta}_{\mathrm{TE}}, \boldsymbol{\theta}_{\mathrm{G}}]$; 
diffusion scheduler (noise schedule $\{\beta_t\}_{t=1}^T$); 
diffusion steps $T$; 
SNR range $[\rho_{\min}, \rho_{\max}]$.
\\
\textbf{Output:} Trained model parameters $\boldsymbol{\theta}^\ast$.

\begin{algorithmic}[1]
    \FOR{$\mathrm{epoch} = 1, \dots, N_{\mathrm{epochs}}$}
        \FOR{each batch $(\mathbf{X}, \mathbf{Y}) \in \mathcal{D}$}
            \STATE $\rho \sim \mathcal{U}[\rho_{\min}, \rho_{\max}]$
            \STATE $\tilde{\mathbf{X}} \gets \sqrt{\rho} \mathbf{X} + \mathbf{N}$, \quad $\mathbf{N} \sim \mathcal{N}(\mathbf{0}, \mathbf{I})$
            \STATE $\mathbf{Z} \gets f_{\boldsymbol{\theta}_\mathrm{TE}}(\tilde{\mathbf{X}})$
            \STATE $t \sim \mathcal{U}[0, T-1]$
            \STATE $\mathbf{X}^{t} \gets \sqrt{\bar{\alpha}_t}\,\mathbf{Y} + \sqrt{1 - \bar{\alpha}_t}\,\boldsymbol{\epsilon}$, \quad $\boldsymbol{\epsilon} \sim \mathcal{N}(\mathbf{0}, \mathbf{I})$
            \STATE $\mathbf{Y}_{\mathrm{diff}} \gets \mathrm{Concat}(\mathbf{X}^{t}, \mathbf{Z})$
            \STATE $\hat{\mathbf{Y}} \gets f_{\boldsymbol{\theta}_{\mathrm{G}}}(\mathbf{Y}_{\mathrm{diff}}, t)$
            \STATE $\mathcal{L}(\theta) \gets \mathrm{Loss}(\hat{\mathbf{Y}}, \mathbf{Y})$
            \STATE Update $\boldsymbol{\theta}$ using gradients $\nabla_{\boldsymbol{\theta}}\mathcal{L}(\boldsymbol{\theta})$
        \ENDFOR
    \ENDFOR
\STATE \textbf{Return} $\boldsymbol{\theta}^\ast$    
\end{algorithmic}
\end{algorithm}

We apply two different inference schemes: the first follows an AR approach, while the second adopts a seq2seq prediction framework. In the following sections, we elaborate on these prediction schemes and summarize the overall methodology in algorithmic formulations.
The overall training procedure of our diffusion-based CSI predictor, which leverages latent representations from the temporal encoder, is outlined in Algorithm~\ref{alg:training_procedure}. Note that the training process is identical for both AR and seq2seq schemes, except for the output target: in the AR scenario, the model is trained to generate the next CSI frame, whereas in the seq2seq case, it is trained to produce a sequence of CSI samples over a fixed prediction horizon.

\subsection{AR}

\begin{figure}
    \centering
    \includegraphics[height=6cm, width=8.5cm]{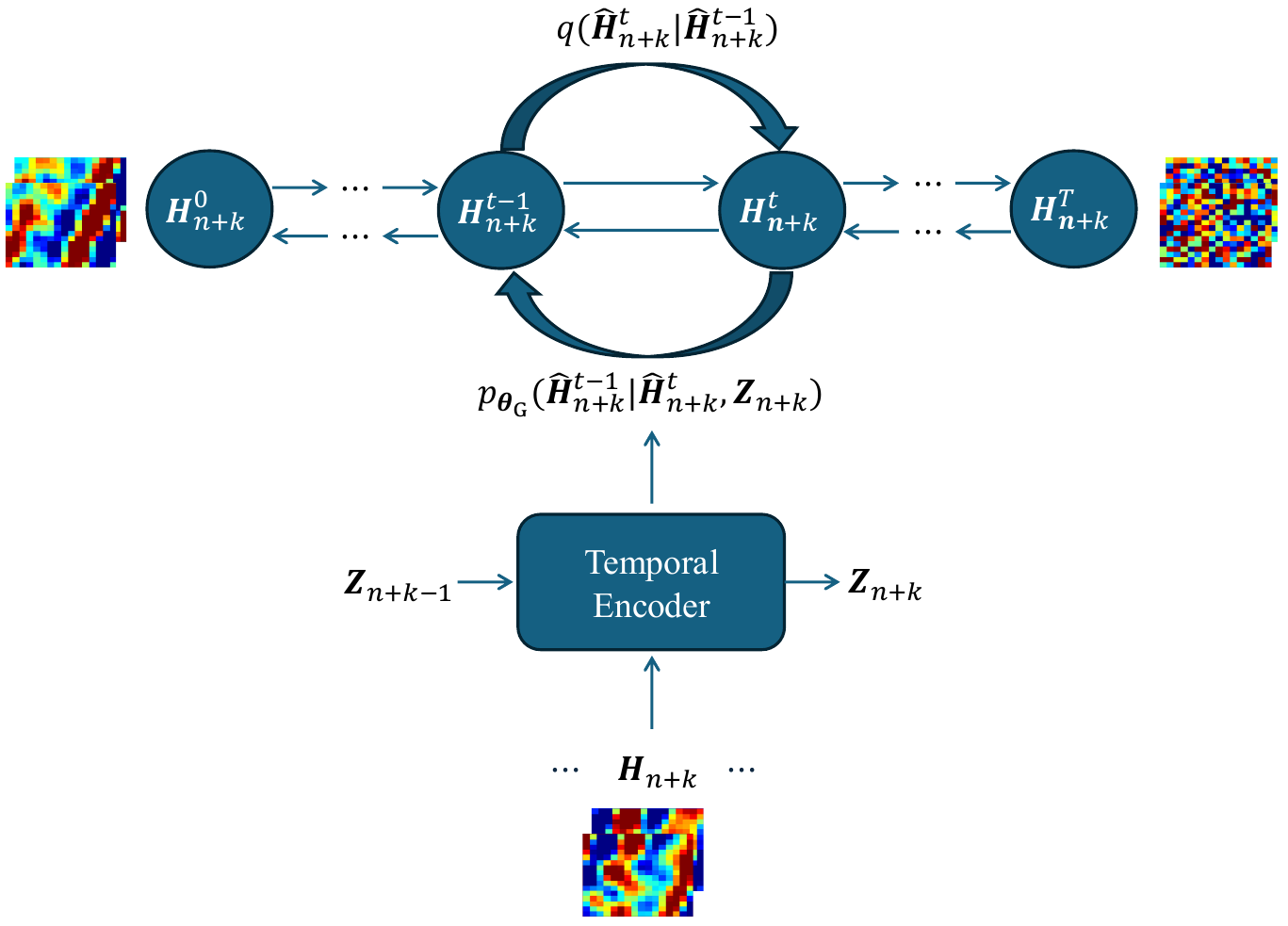}
    \caption{CSI prediction with a diffusion model conditioned on the temporal encoder’s latent representation using an AR inference strategy.}
    \label{fig: AR_TE_DM}
\end{figure}

AR inference is a sequential prediction strategy in which each future time step is generated conditioned on previously observed or predicted steps. At each step, the model predicts the next CSI frame and recursively feeds this prediction back to forecast subsequent frames until the desired horizon is reached. 
This strategy provides a flexible prediction framework, where a single trained model can operate with arbitrary context lengths and prediction horizons. Such flexibility is particularly suitable for sequential channel modeling, as the predictor continuously adapts to the evolving CSI sequence.
The overall AR CSI prediction framework is illustrated in Fig.~\ref{fig: AR_TE_DM}.

\paragraph{Temporal Encoding}

Learning the dynamics of CSI data is a crucial component of the CSI prediction problem. To capture the complex nature of a time-varying wireless channel, we employ a learning model to extract latent features along the temporal dimension.
At the $k^{\text{th}}$ prediction step, the encoder input is
\begin{equation}
    \mathbf{H}_{n+k} = \{\mathbf{H}_{n-N_{\mathrm{p}}+1:n}, \hat{\mathbf{H}}_{n+1:n+k-1} \},
\end{equation}
where $\mathbf{H}_{n-N_{\mathrm{p}}+1:n}$ denotes the observed CSI frames and $\hat{\mathbf{H}}_{n+1:n+k-1}$ represents previously predicted frames.
The temporal encoder, parameterized by $\boldsymbol{\theta}_{\mathrm{TE}}$, produces a step-dependent latent representation
\begin{equation}
     \mathbf{Z}_{n+k} = f_{\boldsymbol{\theta}_{\mathrm{TE}}}(\mathbf{H}_{n+k}).
\end{equation}

The latent feature $\mathbf{Z}_{n+k}$ captures evolving temporal correlations, temporal coherence, Doppler-induced continuity, and multipath dynamics. Since the encoder is updated at every prediction step, the representation adapts to the progressively generated CSI sequence. This dynamic conditioning signal guides the diffusion generator to produce the next CSI frame at time $n+k$.

\paragraph{Generator}

To generate future CSI conditioned on the extracted latent representation, we employ a diffusion model. In the forward process, Gaussian noise is progressively added according to the Markov transition in~\eqref{forward_diffusion_transition}. 

In the reverse process, a neural network parameterized by $\boldsymbol{\theta}_{\mathrm{G}}$ is trained to approximate the conditional distribution
\begin{equation}
    f_{\boldsymbol{\theta}_\mathrm{G}}(\hat{\mathbf{H}}_{n+k}^t, \mathbf{Z}_{n+k}) 
    \approx 
    p_{\boldsymbol{\theta}_\mathrm{G}}(\hat{\mathbf{H}}_{n+k}^{t-1} 
    \mid 
    \hat{\mathbf{H}}_{n+k}^t, \mathbf{Z}_{n+k}),
\end{equation}
thereby enabling iterative denoising conditioned on the temporal latent feature. After $T$ sampling steps, the predicted CSI frame is obtained as
\begin{align}
    \hat{\mathbf{H}}_{n+k} = \hat{\mathbf{H}}_{n+k}^{0}.
\end{align}

\subsubsection*{Inference}

During inference, each generated CSI frame is appended to the input sequence and fed back into the temporal encoder to predict the next step. This recursive process continues until the desired prediction horizon is reached, as summarized in Algorithm~\ref{alg:inference_AR}.

\begin{algorithm}[t]
\caption{AR Inference}
\label{alg:inference_AR}
\textbf{Input:} Historical CSI sequence $\mathbf{H}_\mathrm{p}$; 
trained parameters $\boldsymbol{\theta}^\ast = [\boldsymbol{\theta}_\mathrm{TE}^\ast, \boldsymbol{\theta}_\mathrm{G}^\ast]$; 
diffusion scheduler with noise schedule $\{\beta_t\}_{t=1}^{T}$; 
diffusion steps $T$; 
prediction horizon $N_\mathrm{f}$.
\\
\textbf{Output:} Predicted CSI sequence $\{\hat{\mathbf{H}}_n\}_{n=1}^{N_\mathrm{f}}$.

\begin{algorithmic}[1]
    \STATE Initialize $\mathbf{H}_{c} \gets \mathbf{H}_\mathrm{p}$
    \FOR{$n = 1$ \TO $N_\mathrm{f}$}
        \STATE $\mathbf{Z} \gets f_{\boldsymbol{\theta}^\ast_\mathrm{TE}}(\mathbf{H}_\mathrm{c})$
        \STATE Initialize $\mathbf{H}^{T} \sim \mathcal{N}(\mathbf{0}, \mathbf{I})$
        \FOR{$t = T$ \TO $1$}
            \STATE $\hat{\mathbf{H}}^0 \gets f_{\boldsymbol{\theta}^\ast_\mathrm{G}}(\mathrm{Concat}(\mathbf{H}^{t}, \mathbf{Z}), t)$
            \STATE $\boldsymbol{\epsilon}_{\boldsymbol{\theta}}(\mathbf{H}^t, t) = \frac{\mathbf{H}^t - \sqrt{\bar{\alpha}_t}\,\hat{\mathbf{H}}^0_{\boldsymbol{\theta}}(\mathbf{H}^t, t)}{\sqrt{1 - \bar{\alpha}_t}}$
            \STATE $\mathbf{H}^{t-1} \gets \sqrt{\bar{\alpha}_{t-1}} \, \hat{\mathbf{H}}^0 + \sqrt{1 - \bar{\alpha}_{t-1}} \, \boldsymbol{\epsilon}_{\boldsymbol{\theta}}(\mathbf{H}^t, t)$
        \ENDFOR
        \STATE $\hat{\mathbf{H}}_n \gets \hat{\mathbf{H}}^0$
        \STATE $\tilde{\mathbf{H}}_n \gets \hat{\mathbf{H}}_n + \mathbf{N}$, \quad $\mathbf{N} \sim \mathcal{N}(\mathbf{0}, \sigma^2\mathbf{I})$
        \STATE $\mathbf{H}_\mathrm{c} \gets \mathrm{Concat}(\mathbf{H}_\mathrm{c}, \tilde{\mathbf{H}}_n)$
    \ENDFOR
    \STATE \textbf{Return} $\{\hat{\mathbf{H}}_n\}_{n=1}^{N_\mathrm{f}}$
\end{algorithmic}
\end{algorithm}

\subsection{Seq2seq}
Seq2seq inference is a parallel prediction strategy in which the entire sequence of future steps is generated simultaneously rather than recursively. Instead of predicting one step at a time and feeding it back into the model, a seq2seq approach maps the input sequence directly to multiple outputs in a single forward pass. 

\paragraph{Temporal Encoding}  
In the seq2seq prediction framework, the temporal encoder processes a fixed-length CSI sequence and extracts a latent representation,  
\begin{equation}
    \mathbf{Z} = f_{\boldsymbol{\theta}_\mathrm{TE}}(\mathbf{H}_{\mathrm{p}}).
\end{equation}

The latent feature $\mathbf{Z}$ encodes a global summary of the channel evolution over the observation window, capturing temporal coherence, Doppler continuity, and multipath dynamics embedded in the historical CSI sequence. It conditions the diffusion generator to jointly generate multiple future CSI frames in parallel.

\paragraph{Generator}  
Conditioned on this latent feature, the generator produces the entire sequence of future CSI samples over a fixed prediction horizon in parallel using a diffusion process. The generator then learns to approximate the reverse process by modeling the conditional distribution  
\begin{equation}
    f_{\boldsymbol{\theta}_\mathrm{G}}(\hat{\mathbf{H}}^{t} , \mathbf{Z}) \approx
    p_{\boldsymbol{\theta}_\mathrm{G}}(\hat{\mathbf{H}}^{t-1} \mid \hat{\mathbf{H}}^t, \mathbf{Z}),
\end{equation}
and after $T$ sampling steps, the predicted CSI sample is
\begin{align}
    \hat{\mathbf{H}}_{\mathrm{f}}=\hat{\mathbf{H}}^{0}.
\end{align}

This design eliminates the sequential dependency of AR inference, reducing inference time and computational overhead, especially for long prediction horizons. By predicting all future steps in parallel, seq2seq inference also avoids the problem of error accumulation, since early mistakes do not propagate through the sequence. However, this efficiency comes at the cost of reduced flexibility as a seq2seq model can be trained for a fixed context length and prediction horizon, and may struggle to capture fine-grained temporal dependencies compared to AR methods.

\subsubsection*{Inference}
During inference, the trained temporal encoder processes the historical CSI samples and encodes them into a temporal latent representation. This representation is concatenated with standard Gaussian noise and passed to the diffusion generator, where the reverse diffusion process is carried out using the DDIM scheduler as outlined in Algorithm~\ref{alg:inference_seq2seq}. 

It is worth noting that various temporal encoders (e.g., RNNs, Transformers) and backbone models for diffusion (e.g., U-Net, DiT) can be employed. In Section~\ref{sec4}, we examine our diffusion-based CSI prediction framework under different architectural choices.

\begin{algorithm}[t]
\caption{Seq2seq Inference}
\label{alg:inference_seq2seq}
\textbf{Input:} Historical CSI sequence $\mathbf{H}_\mathrm{p}$ of length $N_\mathrm{p}$; 
trained parameters $\boldsymbol{\theta}^\ast = [\boldsymbol{\theta}_\mathrm{TE}^\ast, \boldsymbol{\theta}_\mathrm{G}^\ast]$; 
diffusion scheduler with noise schedule $\{\beta_t\}_{t=1}^{T}$; 
diffusion steps $T$.
\\
\textbf{Output:} Predicted CSI sequence $\{\hat{\mathbf{H}}_\mathrm{f}\}$.

\begin{algorithmic}[1]
    \STATE $\mathbf{Z} \gets f_{\boldsymbol{\theta}^\ast_\mathrm{TE}}(\mathbf{H}_\mathrm{p})$
    \STATE Initialize $\mathbf{H}^{T} \sim \mathcal{N}(\mathbf{0}, \mathbf{I})$
    \FOR{$t = T$ \TO $1$}
        \STATE $\hat{\mathbf{H}}^0 \gets f_{\boldsymbol{\theta}^\ast_\mathrm{G}}(\mathrm{Concat}(\mathbf{H}^{t}, \mathbf{Z}), t)$
        \STATE $\boldsymbol{\epsilon}_{\boldsymbol{\theta}}(\mathbf{H}^t, t) = \frac{\mathbf{H}^t - \sqrt{\bar{\alpha}_t}\,\hat{\mathbf{H}}^0_{\boldsymbol{\theta}}(\mathbf{H}^t, t)}{\sqrt{1 - \bar{\alpha}_t}}$
        \STATE $\mathbf{H}^{t-1} \gets \sqrt{\bar{\alpha}_{t-1}} \, \hat{\mathbf{H}}^0 + \sqrt{1 - \bar{\alpha}_{t-1}} \, \boldsymbol{\epsilon}_{\boldsymbol{\theta}}(\mathbf{H}^t, t)$
    \ENDFOR
    \STATE $\hat{\mathbf{H}}_\mathrm{f} \gets \hat{\mathbf{H}}^0$
\STATE \textbf{Return} $\hat{\mathbf{H}}_\mathrm{f}$
\end{algorithmic}
\end{algorithm}

\subsection{Unified Seq2seq}  
Employing separate models for temporal encoding and CSI generation provides a robust framework for learning the temporal and spatial characteristics of the dynamic CSI distribution. Training separate models for temporal and spatial dependencies complicates optimization and requires substantial training data. It also increases computational cost, as both networks must be trained and executed for CSI prediction. To address these limitations and enhance practicality in real-world applications, we integrate temporal encoding and generation into a single diffusion model that jointly learns temporal and spatial dependencies from CSI samples. In this formulation, the diffusion model directly approximates the conditional distribution of the CSI dynamics given historical CSI sequences,
\begin{equation}
    p_{\boldsymbol{\theta}_\mathrm{G}}\!\left(\hat{\mathbf{H}}^{t-1} \mid \hat{\mathbf{H}}^{t}, \mathbf{H}_{\mathrm{p}}\right).
\end{equation}

This unified framework reduces model complexity and improves efficiency, though it may be less effective in capturing fine-grained temporal and spatial structures compared to architectures with two dedicated modules.


\section{Simulation Setup}\label{sec4}
In this section, we present a detailed description of the simulation setup used to evaluate the performance of the proposed diffusion-based CSI prediction models. We first introduce the CDL simulation parameters employed to generate the CSI dataset. We then describe the proposed diffusion models with different backbone architectures, including U-Net, DiT, and 3D U-Net, as well as several state-of-the-art CSI prediction baselines for comparison. The training configuration and the techniques adopted during model training are outlined next. Finally, we provide a comprehensive set of numerical results to assess and compare the performance of these models in terms of NMSE across varying prediction horizons and SNR regimes.

\subsection{Dataset Setup}

We generate CSI data using the 3GPP-compliant CDL channel model described in Section~\ref{CDL_model}. A total of 100{,}000 independent CSI samples are created. Each sample is represented as a tensor of shape 
$[100 \times 2 \times N_\mathrm{t} \times N_\mathrm{c}]$, corresponding to 100 OFDM symbols, real and imaginary components, $N_\mathrm{t}$ transmit antennas, and $N_\mathrm{c}$ subcarriers.

Each sample is divided into past CSI $\mathbf{H}_\mathrm{p}$ of length $N_\mathrm{p}$ and future CSI $\mathbf{H}_\mathrm{f}$ of length $N_\mathrm{f}$ to form training pairs $(\mathbf{X}, \mathbf{Y})$, where $\mathbf{X}=\mathbf{H}_\mathrm{p}$ and $\mathbf{Y}=\mathbf{H}_\mathrm{f}$. Unless otherwise specified, we set $N_\mathrm{p}=30$ and $N_\mathrm{f}=10$. Min–max scaling is applied to normalize the dataset to $(0,1)$.
The key simulation parameters are summarized in Table~\ref{tab:simulation_setup}.
\begin{table}[t]
\centering
\caption{Simulation and Dataset Parameters}
\label{tab:simulation_setup}
\renewcommand{\arraystretch}{1.2}
\begin{tabular}{ll}
\toprule
\textbf{Parameter} & \textbf{Value} \\
\midrule
Carrier frequency $f_c$ & 28 GHz \\
Channel model & 3GPP \texttt{CDL-A}–\texttt{E} (randomly selected) \\
Number of BS antennas & 16\\
Resource blocks & 25 RBs \\
Subcarrier spacing & 30 kHz \\
Number of OFDM subcarriers & 16 \\
Symbol duration & $\approx 33.3\,\mu$s \\
User velocity & 30–120 km/h (uniformly sampled) \\
Delay spread & 50–400 ns (uniformly sampled) \\
\bottomrule
\end{tabular}
\end{table}

The cyclic prefix (CP) duration is assumed to exceed the maximum channel delay spread (sampled from $[50,400]$ ns), thereby preventing inter-symbol interference and ensuring circular convolution within each OFDM symbol. Consequently, each subcarrier experiences flat fading, while the delay spread governs the frequency selectivity across subcarriers.

While practical BS deployments typically employ uniform planar arrays (UPAs) to enable three-dimensional beamforming, the considered $16\times 1$ ULA does not limit generality. A UPA structure such as $4\times 4$ MIMO can be equivalently vectorized into a $16\times 1$ spatial-domain representation by stacking the two-dimensional antenna responses into a single spatial vector. Since the proposed framework operates directly on the CSI tensor and treats the spatial dimension in vectorized form, it is agnostic to the underlying array geometry. Therefore, the methodology can be readily extended to practical UPA configurations without modification of the learning architecture.

The proposed CSI prediction framework is duplexing-agnostic. In time-division-duplexing (TDD) operation, downlink CSI is obtained at the BS via uplink pilot-based channel estimation by exploiting channel reciprocity. In frequency-division-duplexing (FDD) operation, downlink CSI is acquired through downlink pilot transmission, channel estimation at the user, and subsequent CSI feedback to the BS. In both cases, the predictor operates on previously estimated CSI realizations.

\subsection{Neural Prediction Models}\label{Prediction Models}
We evaluate the performance of several baseline models and our proposed diffusion-based scheme for CSI prediction. These models are selected to cover a diverse range of architectural paradigms, including RNNs such as GRU and ConvLSTM, efficient Transformer variants such as LinFormer \cite{10965849}, and proposed generative diffusion-based predictors. A brief description of each model is provided below:

\begin{itemize}
    \item \textbf{GRU:} We use a GRU model with two hidden layers, each comprising $128$ channels and employing the $\tanh$ activation function. This serves as a standard sequential baseline for capturing temporal dependencies in CSI prediction.

    \item \textbf{ConvLSTM:} We adopt a single-layer ConvLSTM with a hidden state of $128$ channels and a spatial kernel size of $3 \times 3$, producing latent feature maps in $\mathbb{R}^{128 \times N_\mathrm{t} \times N_\mathrm{c}}$. The output is followed by Group Normalization (single group), Dropout with a rate of $0.2$, a $3 \times 3$ convolutional projection layer with $4$ output channels, and a final $\tanh$ activation. A brief review of ConvLSTM fundamentals is given in Appendix~\ref{appendix: ConvLSTM}.

    \item \textbf{LinFormer:} LinFormer, introduced in \cite{10965849}, is a Transformer variant tailored for CSI prediction. It replaces the computationally expensive self-attention mechanism with a time-aware multi-layer perceptron (TMLP), significantly reducing complexity. The model comprises six encoder blocks, each with a model dimension of $512$ and a feed-forward hidden size of $512$.

    \item \textbf{DiU:} DiU represents a diffusion-based CSI prediction framework employing a U-Net backbone with AR inference. A ConvLSTM acts as the temporal encoder to capture channel dynamics. The U-Net design is detailed in Appendix~\ref{appendix: Unet}.

    \item \textbf{DiU-seq2seq:} This is a seq2seq variant of the diffusion framework using the same U-Net backbone as DiU but predicts multiple future CSI frames in a single forward pass, without an explicit temporal encoder.

    \item \textbf{LinFusion:} LinFusion combines a LinFormer-based temporal encoder with a U-Net diffusion generator under a seq2seq inference scheme. The LinFormer and U-Net components follow the same architecture configurations as in the LinFormer and DiU models, respectively.

    \item \textbf{DiT:} In this variant, we employ a DiT as the denoising backbone, while a ConvLSTM serves as the temporal encoder. Architectural details and parameter settings for DiT are provided in Appendix~\ref{appendix: DiT}.

    \item \textbf{DiU3:} We employ a 3D U-Net as the diffusion backbone with seq2seq inference. The 3D U-Net extends the standard U-Net into three dimensions, jointly modeling spatio-temporal features. The details for the 3D U-Net design used in our simulation are provided in Appendix \ref{appendix:unet3d}.
\end{itemize}

\subsection{Training Parameters}
For training the diffusion model, we employ the DDIM scheduling technique to accelerate the sampling process. Diffusion timesteps $t$ are drawn uniformly from $\{0, \dots, T-1\}$. The input to the diffusion generator is formed by concatenating the noisy CSI at timestep $t$ with the latent features extracted from the temporal encoder. To enhance robustness against varying SNR conditions and to account for channel estimation errors during pilot transmission, we corrupt the training data with additive noise.
The additive white Gaussian noise assumption models the effective channel estimation error under standard pilot-based estimators such as least square (LS) or MMSE.

We adopt the Huber loss~\cite{huber1964} to train both the diffusion model and the temporal encoder, and use the diffusion loss formulated in~\eqref{loss_diff_sample} to predict clean samples. The Huber loss is a robust regression loss that behaves quadratically for small errors and linearly for large errors, combining the benefits of MSE and mean absolute error (MAE). With a threshold parameter $\delta > 0$, it is defined as
\begin{equation}
\mathcal{L}_\delta( y , \hat{y}) =
\begin{cases}
\frac{1}{2}  (y - \hat{y})^2, & \text{if } | y - \hat{y}| \le \delta, \\
\delta \left( | y - \hat{y}| - \frac{\delta}{2} \right), & \text{otherwise}.
\end{cases}
\label{eq:huber_loss}
\end{equation}

We employ the Adam optimizer to jointly update the parameters of the temporal encoder and the diffusion generator, using learning rates of \(1\times 10^{-3}\) and \(1\times 10^{-3}\), respectively. To enhance training stability, we apply an exponential moving average (EMA) with a decay factor of \(0.995\) and an update interval of 10 steps, together with gradient clipping at a maximum norm of 1.0. The model is trained for 500~epochs with a batch size of 512. 

In the implementation, the Huber loss threshold is set to \(\delta = 0.016\), while the SNR values are uniformly sampled from the range \([-20, 20]\)~dB. For the diffusion noise schedule, we adopt the scaled and clipped squared--cosine design proposed in~\cite{nichol2021improveddenoisingdiffusionprobabilistic}, defined as  
\begin{equation}
\bar{\alpha}_t = 
\cos^2\!\left( 
\frac{\tfrac{t}{T} + 0.008}{1.008} \cdot \frac{\pi}{2} 
\right),
\label{eq:squaredcos}
\end{equation}
which leads to the variance schedule  
\begin{equation}
\beta_t = 1 - \frac{\bar{\alpha}_t}{\bar{\alpha}_{t-1}}, \quad t \geq 1.
\end{equation}

To prevent degenerate values, the coefficients are further clipped as  
\begin{equation}
\beta_t = 
\min\!\left( \beta_{\max},\;
\max\!\left( \beta_{\min},\; \beta_t \right) \right),
\label{eq:beta_clip}
\end{equation}
with \(\beta_{\min} = 10^{-4}\), \(\beta_{\max} = 2\times 10^{-2}\), and \(T=2000\) in our experiments. Compared to a linear schedule, the squared--cosine design produces smaller $\beta_t$ values in the early steps, resulting in a slower corruption rate at the start of the forward process, and larger $\beta_t$ near the end, leading to stronger noise injection in later steps.

\subsection{Numerical Results}
We evaluate the performance of the proposed diffusion-based CSI prediction models and baseline approaches introduced in Section~\ref{Prediction Models} under various experimental settings, including different SNR levels, prediction horizons, and user mobility scenarios. The performance of all methods is quantified using the NMSE, defined as
\begin{equation}
\mathrm{NMSE} \triangleq
\mathbb{E}\left[
\frac{\lVert \mathbf{H} - \mathbf{\hat{H}} \rVert^{2}}
{\lVert \mathbf{H} \rVert^{2}}
\right],
\end{equation}
where $\mathbf{H}$ and $\mathbf{\hat{H}}$ denote the ground-truth and predicted channel matrices, respectively.

The SNR is defined as the ratio between the average channel power and the noise power. Assuming unit transmit power, the SNR is given by
\begin{equation}
\mathrm{SNR} \triangleq
\frac{\mathbb{E}\!\left[\|\mathbf{H}\|^2\right]}
{\sigma_n^2},
\end{equation}
where $\sigma_n^2$ is the noise variance. Throughout this paper, SNR values expressed in decibels (dB) are computed using the standard power-based logarithmic representation
\begin{equation}
\mathrm{SNR}_{\mathrm{dB}} = 10\log_{10}(\mathrm{SNR}).
\end{equation}
Similarly, the NMSE values reported in dB are computed as
\begin{equation}
\mathrm{NMSE}_{\mathrm{dB}} = 10\log_{10}(\mathrm{NMSE}),
\end{equation}

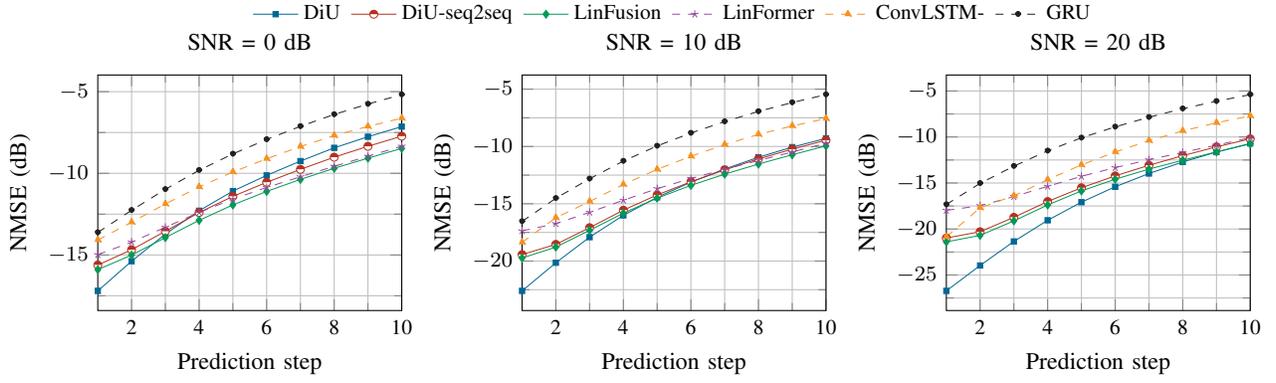
\begin{figure*}[t]
\centering
\begin{tikzpicture}
  \begin{groupplot}[
    group style={group size=3 by 1, horizontal sep=8mm, vertical sep=8mm},
    width=0.35\textwidth,
    height=0.3\textwidth,
    xlabel={Prediction step},
    ylabel={NMSE (dB)},
    ytick distance=5,
    ymajorgrids,
    minor tick num=1,     
    xmin=0,
    xmax=9,
    xtick={1,3,5,7,9},
    xticklabels={2,4,6,8,10}, 
    grid=both,
    cycle list name=mymarkers,
    legend columns=7,
    legend style={/tikz/every even column/.append style={column sep=5pt},
                  draw=none, fill=none, font=\small},
    tick label style={font=\footnotesize},
    label style={font=\normalsize},
    title style={font=\normalsize}
    ]
    \def\stepsize{0,2,3,4,5,6,7,8,9}
    \def\Diffusion{Diffusion}
    \def\LSTM{LSTM}
    \def\thirtyvsten{30vs10}
    \def\GRU{GRU}
    
    \nextgroupplot[title={SNR = 0 dB}, legend to name=ModelsLegendtwo]
      \foreach \M in \modellist {%
        \addplot+ table [x=step, y=\M] {\diverseZero};
        \edef\legendname{\M}%
        \ifx\legendname\Diffusion
          \addlegendentryexpanded{DiU}%
        \else\ifx\legendname\thirtyvsten
          \addlegendentryexpanded{DiU-seq2seq}%
        \else\ifx\legendname\LSTM
          \addlegendentryexpanded{ConvLSTM-}%
        \else\ifx\legendname\GRU
          \addlegendentryexpanded{GRU}%
        \else
          \addlegendentryexpanded{\M}%
        \fi\fi\fi\fi
      }%
    

    \nextgroupplot[title={SNR = 10 dB}, ylabel={}]
      \foreach \M in \modellist {%
        \addplot+ table [x=step, y=\M] {\diverseTen};
      }

    \nextgroupplot[title={SNR = 20 dB}, ylabel={}]
      \foreach \M in \modellist {%
        \addplot+ table [x=step, y=\M] {\diverseTwenty};
      }

  \end{groupplot}

  \node[anchor=south, yshift=20pt]
    at ($(group c1r1.north)!0.5!(group c3r1.north)$)
    {\pgfplotslegendfromname{ModelsLegendtwo}};
\end{tikzpicture}
\caption{NMSE of different CSI prediction models over varying prediction steps at inference SNRs of $0$, $10$, and $20$~dB.}
\label{fig:nmse-prediction_steps}
\end{figure*}

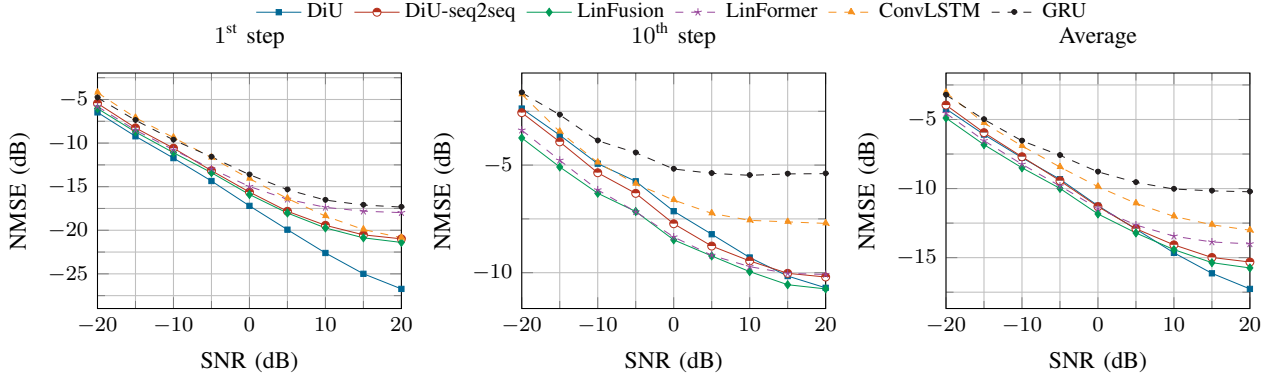
\begin{figure*}[t]
\centering
\begin{tikzpicture}
  \begin{groupplot}[
    group style={group size=3 by 1, horizontal sep=8mm, vertical sep=6mm},
    width=0.35\textwidth,
    height=0.3\textwidth,
    xlabel={SNR (dB)},
    ylabel={NMSE (dB)},
    ytick distance=5,
    ymajorgrids,
    minor tick num=1,
    xmin=-20,
    xmax=20,
    grid=both,
    cycle list name=mymarkers,
    legend columns=7,
    legend style={/tikz/every even column/.append style={column sep=4pt},
                  draw=none, fill=none, font=\small},
    tick label style={font=\footnotesize},
    label style={font=\normalsize},
    title style={font=\normalsize}
  ]

    \def\Diffusion{Diffusion}
    \def\DIT{DIT}
    \def\UnetThreeD{Unet3D}
    \def\LSTM{LSTM}
    \def\thirtyvsten{30vs10}
    \def\GRU{GRU}

    \nextgroupplot[title={\(1^\text{st}\) step}, legend to name=ModelsLegend]
      \foreach \M in \modellist {%
        \edef\col{\M_first}%
        \addplot+ table [x=snr_db, y=\col] {\diversesummary};
        \edef\legendname{\M}%
        \ifx\legendname\Diffusion
          \addlegendentryexpanded{DiU}%
        \else\ifx\legendname\thirtyvsten
          \addlegendentryexpanded{DiU-seq2seq}%
        \else\ifx\legendname\LSTM
          \addlegendentryexpanded{ConvLSTM}%
        \else\ifx\legendname\GRU
          \addlegendentryexpanded{GRU}%
        \else
          \addlegendentryexpanded{\M}%
        \fi\fi\fi\fi
      }%

    \nextgroupplot[title={\(10^\text{th}\) step}, ylabel={}]
      \foreach \M in \modellist {%
        \edef\col{\M_last}%
        \addplot+ table [x=snr_db, y=\col] {\diversesummary};
      }

    \nextgroupplot[title={Average}, ylabel={}]
      \foreach \M in \modellist {%
        \edef\col{\M_avg}%
        \addplot+ table [x=snr_db, y=\col] {\diversesummary};
      }

  \end{groupplot}

  \node[anchor=south, yshift=20pt]
    at ($(group c1r1.north)!0.5!(group c3r1.north)$)
    {\pgfplotslegendfromname{ModelsLegend}};
\end{tikzpicture}
\caption{NMSE performance of CSI prediction models versus inference SNR. Results are reported for the first prediction step, the last prediction step, and the average NMSE across all prediction steps.}
\label{fig:nmse-snr}
\end{figure*}

Fig.~\ref{fig:nmse-prediction_steps} illustrates the prediction accuracy of different models across varying prediction steps under three inference SNR regimes: $0$, $10$, and $20$~dB. The results show that the proposed diffusion models consistently outperform state-of-the-art benchmarks such as GRU, ConvLSTM, and LinFormer. The performance gain is most pronounced at higher SNRs and shorter prediction horizons. For example, at an inference SNR of $20$~dB, the diffusion model with AR inference achieves improvements of more than $5$~dB and $8$~dB in NMSE compared to ConvLSTM and LinFormer, respectively. When comparing the three diffusion-based predictors, the AR variant noticeably outperforms the seq2seq variants. In contrast, the seq2seq diffusion models provide slightly better NMSE performance at lower SNRs and for longer prediction horizons, primarily because of error propagation over time in AR inference. Furthermore, incorporating diffusion into the Linformer architecture yields an NMSE improvement of approximately $4$ dB compared to the Linformer baseline. While LinFusion outperforms Linformer alone, the diffusion model using a ConvLSTM temporal encoder and a U-Net backbone consistently achieves the lowest NMSE, particularly for short prediction horizons.

Fig.~\ref{fig:nmse-snr} presents the NMSE performance of different CSI prediction models as a function of inference SNR, evaluated across three prediction horizons: $1^\text{st}$-step prediction, $10^\text{th}$-step prediction, and the average performance over $10$ prediction steps. As anticipated, higher SNR levels enhance the performance of all models, with our diffusion-based schemes consistently surpassing the baselines. Specifically, for the average NMSE across all steps, the diffusion model with a U-Net backbone and AR inference achieves the best overall performance. For first-step prediction, DiU with AR inference yields up to $5$--$8$~dB NMSE gains compared to GRU, ConvLSTM, and LinFormer, particularly at high SNRs. At longer prediction horizons, the seq2seq diffusion models surpass AR inference, though at the expense of reduced accuracy for short horizons and lower flexibility in handling variable context lengths and prediction horizons.

\begin{figure*}[t]
\centering
\begin{tikzpicture}
  \begin{groupplot}[
    group style={group size=3 by 1, horizontal sep=8mm, vertical sep=6mm},
    width=0.35\textwidth,
    height=0.3\textwidth,
    xlabel={Prediction step},
    ylabel={NMSE (dB)},
    ytick distance=5,
    ymajorgrids,
    minor tick num=1,
    xmin=0, xmax=9,
    xtick={1,3,5,7,9},
    xticklabels={2,4,6,8,10}, 
    grid=both,
    cycle list name=mymarkers,
    legend columns=7,
    legend style={/tikz/every even column/.append style={column sep=4pt},
                  draw=none, fill=none, font=\small},
    tick label style={font=\footnotesize},
    label style={font=\normalsize},
    title style={font=\normalsize},
  ]

    \def\Diffusion{Diffusion}
    \def\DIT{DIT}
    \def\UnetThreeD{Unet3D}
    \def\LSTM{LSTM}
    \def\thirtyvsten{30vs10}
    \def\GRU{GRU}

    \nextgroupplot[title={30 km/h}, legend to name=ModelsLegendthree]
      \foreach \M in \modellist {%
        \addplot+ table [x=step, y=\M] {\dataThirty};
        \edef\legendname{\M}%
        \ifx\legendname\Diffusion
          \addlegendentryexpanded{DiU}%
        \else\ifx\legendname\thirtyvsten
          \addlegendentryexpanded{DiU-seq2seq}%
        \else\ifx\legendname\LSTM
          \addlegendentryexpanded{ConvLSTM}%
        \else\ifx\legendname\GRU
          \addlegendentryexpanded{GRU}%
        \else
          \addlegendentryexpanded{\M}%
        \fi\fi\fi\fi
      }%

    \nextgroupplot[title={60 km/h}, ylabel={}]
      \foreach \M in \modellist {%
        \addplot+ table [x=step, y=\M] {\dataSixty};
      }

    \nextgroupplot[title={120 km/h}, ylabel={}]
      \foreach \M in \modellist {%
        \addplot+ table [x=step, y=\M] {\dataOneTwenty};
      }

  \end{groupplot}

  \node[anchor=south, yshift=20pt]
    at ($(group c1r1.north)!0.5!(group c3r1.north)$)
    {\pgfplotslegendfromname{ModelsLegendthree}};
\end{tikzpicture}
\caption{NMSE performance of CSI prediction models as a function of prediction step, evaluated on CSI samples with user mobilities of $30$~km/h, $60$~km/h, and $120$~km/h.}
\label{fig:nmse-velocity}
\end{figure*}
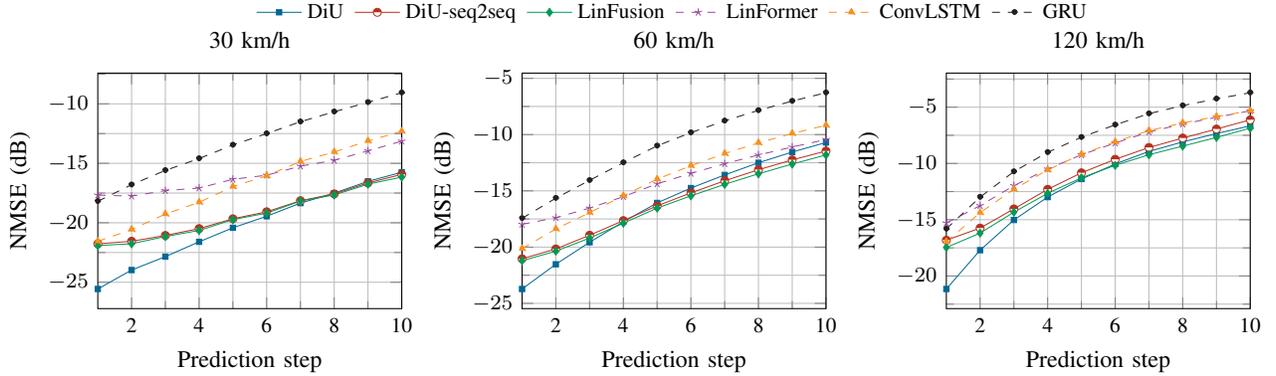

We evaluate the prediction accuracy of different models under fixed user mobility levels of $30$ km/h, $60$ km/h, and $120$ km/h, as shown in Fig.~\ref{fig:nmse-velocity}. As expected, the prediction performance degrades with increasing mobility due to faster channel dynamics and weaker temporal correlations. Nevertheless, the proposed diffusion-based predictors consistently outperform all baseline models across all velocity regimes. The NMSE gains of AR-based diffusion are particularly pronounced at lower user speeds (e.g., $30$ km/h), where stronger temporal correlations can be exploited. In this regime and for short prediction horizons, DiU with AR inference achieves up to a $5$~dB improvement over both DiU-seq2seq and LinFusion. However, this performance gap gradually diminishes as the prediction horizon extends and the user velocity increases.

\begin{figure*}[t]
\centering
\input{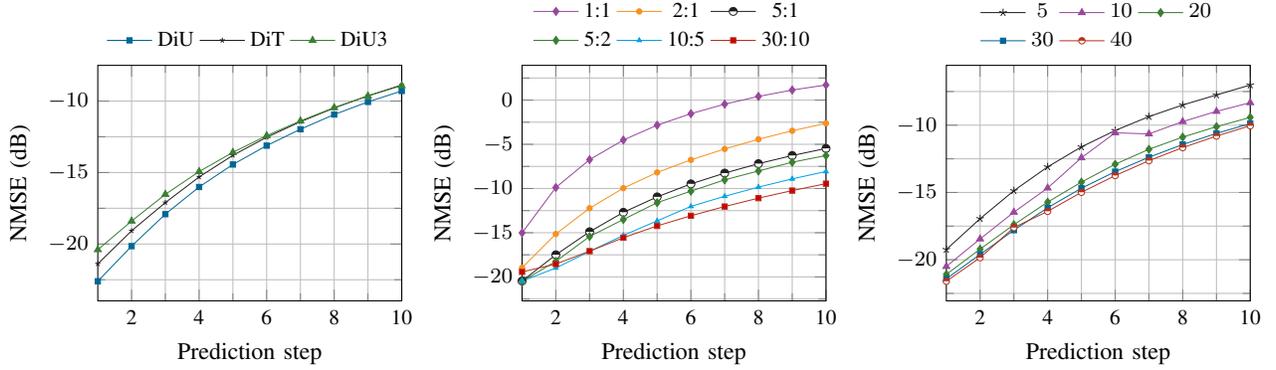}
\caption{NMSE performance versus prediction step at $\mathrm{SNR}=10$~dB. The left panel compares DiU, DiT, and DiU3; the middle panel shows different seq2seq variants; and the right panel presents DiU with varying context lengths.}
\label{fig:linfusion-single-shot-context}
\end{figure*}

Fig.~\ref{fig:linfusion-single-shot-context} compares the CSI prediction performance of different backbone architectures and context lengths across varying prediction horizons. The left plot evaluates our proposed diffusion-based predictors with U-Net, DiT, and 3D U-Net backbones. These results demonstrate the flexibility of the diffusion framework, as different architectures can be seamlessly integrated into the generator.
For DiU and DiT, the same temporal encoder (ConvLSTM) and AR inference scheme are used, and only the diffusion backbone is changed. For the 3D U-Net variant, temporal and spatial dependencies are jointly modeled through 3D convolutions, and therefore no explicit temporal encoder is used.
While all models achieve comparable performance, the U-Net backbone consistently achieves the lowest NMSE. Moreover, although the 3D U-Net variant applies convolutional operations over the 3D volume of CSI data and has significantly higher model capacity, it does not outperform the 2D U-Net variant combined with an explicit temporal encoder and AR inference. This underscores the effectiveness of our proposed diffusion-based CSI prediction framework with an explicit temporal encoder.

The middle plot compares different variants of the seq2seq inference strategy for diffusion-based CSI prediction using DiU. We evaluate multiple configurations, including $1{:}1$, $2{:}1$, $5{:}1$, $5{:}2$, $10{:}5$, and $30{:}10$, where an $N{:}M$ setup denotes predicting $M$ future CSI frames from $N$ historical observations.
In this setup, the prediction framework (DiU) is kept fixed, and only the inference configuration is varied. In particular, $30{:}10$ corresponds to pure seq2seq inference, whereas $1{:}1$, $2{:}1$, and $5{:}1$ correspond to purely AR inference. Intermediate configurations such as $5{:}2$ and $10{:}5$ combine seq2seq prediction with AR recurrence.
The results show that AR-based schemes outperform seq2seq inference for short prediction horizons, even with shorter context lengths, while pure seq2seq inference achieves better performance for longer prediction horizons.

Finally, the right plot highlights the impact of varying the input context length $N_\mathrm{p}$ on prediction accuracy of DiU with AR inference. Increasing the context length from $N_\mathrm{p}= 5$ to $N_\mathrm{p} = 40$ steadily improves NMSE performance, demonstrating that the diffusion model benefits from larger temporal contexts. However, the performance gain begins to saturate beyond $N_\mathrm{p} = 30$, indicating diminishing returns when excessively long histories are used.


\begin{figure*}[t]
\centering
\begin{tikzpicture}
  \begin{groupplot}[
    group style={group size=3 by 1, horizontal sep=8mm, vertical sep=8mm},
    width=0.35\textwidth,
    height=0.3\textwidth,
    xlabel={Prediction step},
    ylabel={SE (bits/s/Hz)},
    ymajorgrids,
    minor tick num=1,     
    xmin=0,
    xmax=9,
    xtick={1,3,5,7,9},
    xticklabels={2,4,6,8,10}, 
    grid=both,
    cycle list name=mymarkers,
    legend columns=7,
    legend style={/tikz/every even column/.append style={column sep=5pt},
                  draw=none, fill=none, font=\small},
    tick label style={font=\footnotesize},
    label style={font=\normalsize},
    title style={font=\normalsize}
    ]
    \def\stepsize{0,2,3,4,5,6,7,8,9}
    \def\Diffusion{Diu}
    \def\LSTM{Conv-LSTM}
    \def\thirtyvsten{30vs10}
    \def\GRU{GRU}
    
    \nextgroupplot[title={SNR = 0 dB}, legend to name=ModelsLegendtwo]
      \foreach \M in \modellistSE {%
        \addplot+ table [x=step, y=\M] {\diverseZeroSe};
        \edef\legendname{\M}%
        \ifx\legendname\Diffusion
          \addlegendentryexpanded{DiU}%
        \else\ifx\legendname\thirtyvsten
          \addlegendentryexpanded{DiU-seq2seq}%
        \else\ifx\legendname\LSTM
          \addlegendentryexpanded{ConvLSTM}%
        \else\ifx\legendname\GRU
          \addlegendentryexpanded{GRU}%
        \else
          \addlegendentryexpanded{\M}%
        \fi\fi\fi\fi
      }%
    

    \nextgroupplot[title={SNR = 10 dB}, ylabel={}]
      \foreach \M in \modellistSE {%
        \addplot+ table [x=step, y=\M] {\diverseTenSe};
      }%

    \nextgroupplot[title={SNR = 20 dB}, ylabel={}]
      \foreach \M in \modellistSE {%
        \addplot+ table [x=step, y=\M] {\diverseTwentySe};
      }

  \end{groupplot}

  \node[anchor=south, yshift=20pt]
    at ($(group c1r1.north)!0.5!(group c3r1.north)$)
    {\pgfplotslegendfromname{ModelsLegendtwo}};
\end{tikzpicture}
\caption{SE of CSI prediction models as a function of the prediction horizon at inference SNRs of $0$, $10$, and $20$~dB.}
\label{fig:se-snr}
\end{figure*}
\begin{figure*}[t]
\centering
\begin{minipage}{0.45\textwidth}
\centering
\pgfplotstableread[col sep=comma]{\pathOneTwenty/nmse_steps_by_sampling.csv}\datatable

\begin{tikzpicture}

  \begin{axis}[
    width=\textwidth,
    height=0.7\textwidth,
    xlabel={Prediction step},
    ylabel={NMSE (dB)},
    xmin=0, xmax=9,
    ymin=-25, ymax=-5, 
    xtick={1,3,5,7,9},
    xticklabels={2,4,6,8,10},
    ytick distance=5,
    minor tick num=1,
    grid=both,
    ymajorgrids,
    tick label style={font=\footnotesize},
    label style={font=\normalsize},
    legend style={
      at={(0.5,1.02)},
      anchor=south,
      legend columns=3,
      yshift=5pt,
      /tikz/every even column/.append style={column sep=5pt},
      font=\small,
      draw=none,
      fill=none
    },
    cycle list name=mymarkers2,
  ]
    \def\samplist{2,3,4,5,10,100}
    \foreach \s in \samplist {%
      \addplot+ table [x=step, y=\s] {\datatable};
      \addlegendentryexpanded{$T=\s$}
    }

    \draw[gray, dashed]
      (axis cs:5.8,-9.9) rectangle (axis cs:6.3,-8.5);

  \end{axis}

  \begin{axis}[
    width=3.6cm,
    height=3.2cm,
    at={(4.3cm,0.55cm)},
    anchor=south west,
    xmin=5.8, xmax=6.3,
    ymin=-9.9, ymax=-8.5,
    tick label style={font=\scriptsize},
    label style={font=\scriptsize},
    grid=both,
    minor tick num=1,
    cycle list name=mymarkers2,
    legend style={draw=none, fill=none},
  ]
    \def\samplist{2,3,4,5,10,100}
    \foreach \s in \samplist {%
      \addplot+ table [x=step, y=\s] {\datatable};
    }
  \end{axis}

  \draw[gray, dashed]
    (4.7cm,3.cm) -- (5.3cm,2.2cm);

\end{tikzpicture}
\end{minipage}
\hfill
\begin{minipage}{0.5\textwidth}
\centering
\vspace{5pt}
\includegraphics[width=\textwidth]{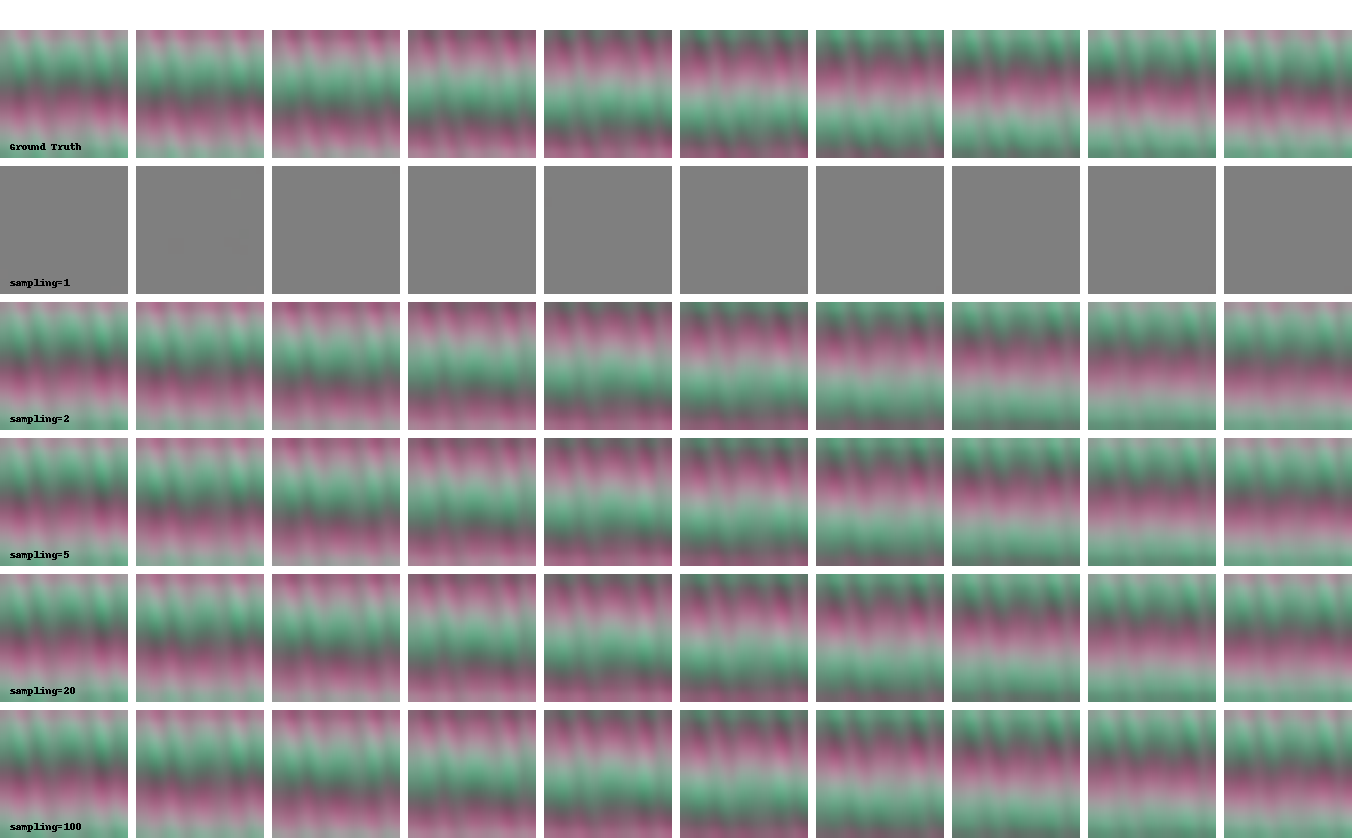}
\end{minipage}

\caption{Left: NMSE performance versus prediction step at $10$~dB SNR for different sampling steps. Right: qualitative channel visualization for a single sequence, where rows correspond to the ground truth and predictions for each sampling step, and columns indicate the prediction steps.}
\label{fig:sampling_steps}
\end{figure*}

To evaluate the performance of various prediction models in downstream tasks, we report the spectral efficiency (SE) achieved by different prediction models under inference SNRs of 0, 10, and 20 dB in Fig. \ref{fig:se-snr}. The SE is computed using a simple maximum-ratio (MR) linear precoding scheme. Consistent with the NMSE trends in Fig. \ref{fig:nmse-snr}, diffusion-based CSI predictors provide higher SE compared to conventional baseline models, particularly at higher SNRs and shorter prediction horizons. These results further highlight the direct relationship between prediction accuracy and system-level performance, confirming that improved NMSE translates into higher spectral efficiency.

Fig.~\ref{fig:sampling_steps} illustrates the impact of different sampling steps on the performance of the proposed diffusion-based 
CSI prediction model at an inference $\mathrm{SNR} = 10~\mathrm{dB}$. The left plot shows the NMSE as a function of prediction step 
for different sampling schedules, while the right panel provides a qualitative visualization of reconstructed CSI frames for one 
representative sequence. Rows correspond to different sampling step settings, and columns represent future prediction steps.

The results demonstrate that the proposed model achieves competitive performance even with a small number of sampling steps. 
In particular, using as few as $3$ sampling steps achieves nearly the same NMSE as the $100$-step schedule, while significantly 
reducing inference time. Beyond $5$ steps, the performance improvements become marginal, indicating that the proposed method 
can operate efficiently with very few diffusion iterations. The qualitative reconstructions on the right further confirm that 
high-fidelity predictions can be achieved even under low sampling budgets.

\begin{figure*}[t]
\centering
\begin{minipage}{0.45\textwidth}
\centering
\begin{tikzpicture}
  \begin{axis}[
    width=\textwidth,
    height=0.7\textwidth,
    xlabel={Prediction step},
    ylabel={NMSE (dB)},
    xmin=1, xmax=10,
    xtick={2,4,6,8,10},
    ytick distance=5,
    ymajorgrids,
    minor tick num=1,
    grid=both,
    cycle list name=mymarkers,
    legend columns=3,
    legend style={
      at={(0.5,1.02)},
      anchor=south,
      yshift=5pt,
      /tikz/every even column/.append style={column sep=5pt},
      draw=none,
      fill=none,
      font=\small
    },
    tick label style={font=\footnotesize},
    label style={font=\normalsize},
    title style={font=\normalsize},
  ]

    \def\Diffusion{Diffusion}
    \def\DIT{DIT}
    \def\UnetThreeD{Unet3D}
    \def\LSTM{LSTM}

    \def\modellist{{Diffusion},{DIT},{Unet3D},{LinFormer},{LSTM},{GRU}}

    \foreach \M in \modellist {%
      \addplot+ table [x=Step, y=\M] {\threeGHz};
      \edef\legendname{\M}%
      \ifx\legendname\Diffusion
        \addlegendentryexpanded{DiU}%
      \else\ifx\legendname\DIT
        \addlegendentryexpanded{DiT}%
      \else\ifx\legendname\UnetThreeD
        \addlegendentryexpanded{DiU3}%
      \else\ifx\legendname\LSTM
        \addlegendentryexpanded{ConvLSTM}%
      \else
        \addlegendentryexpanded{\M}%
      \fi\fi\fi\fi
    }

  \end{axis}
\end{tikzpicture}
\end{minipage}
\hfill
\begin{minipage}{0.5\textwidth}
\centering
\vspace{5pt}
\includegraphics[width=\textwidth, height = .7\textwidth]{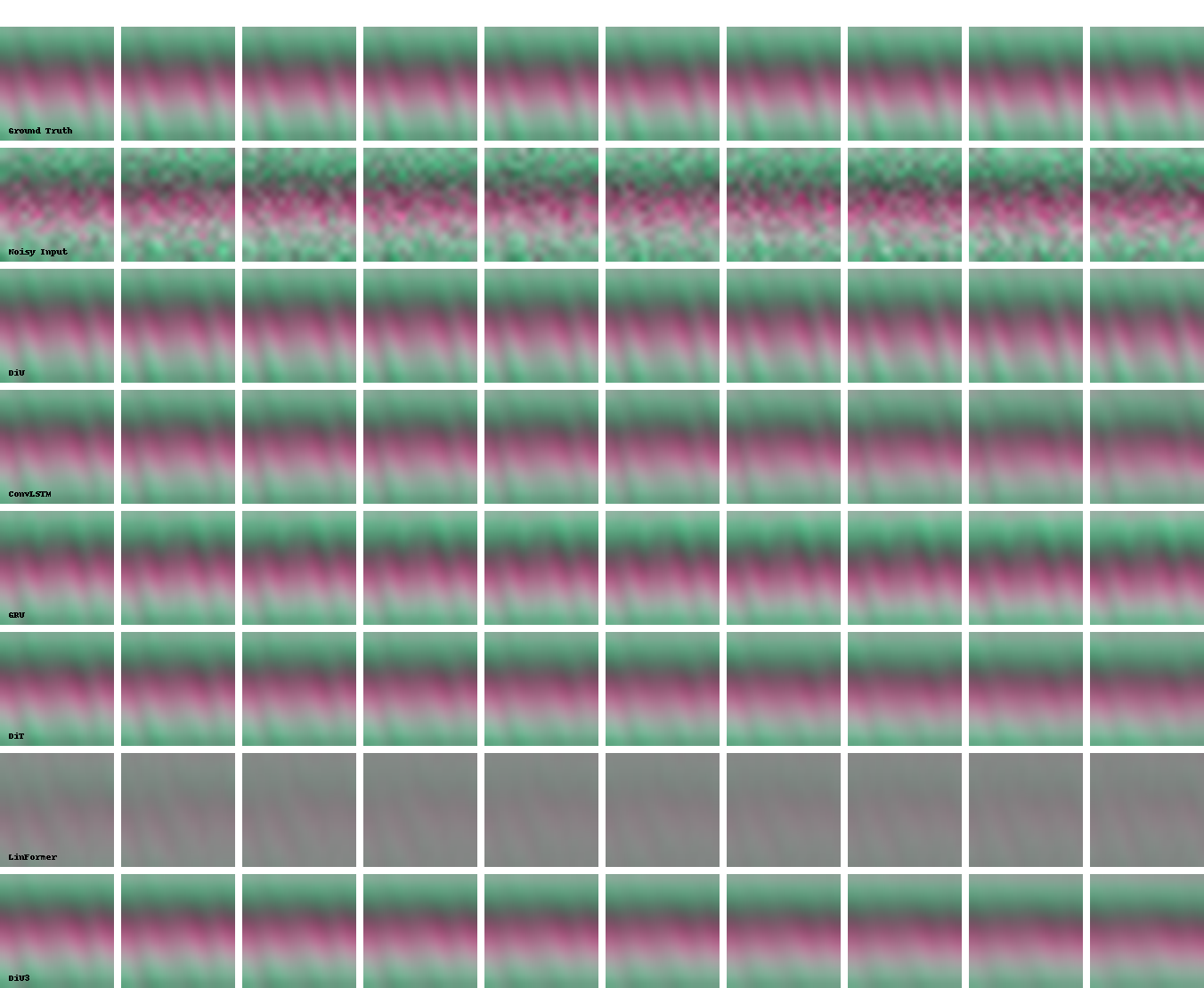}
\end{minipage}

\caption{Left: NMSE performance versus prediction step, evaluated at $f_\mathrm{c}=3$~GHz. Right: qualitative channel visualization for a single sequence, where rows correspond to the ground truth and different prediction schemes, and columns represent the prediction steps.}
\label{fig:fc_3GHz}
\end{figure*}

Fig.~\ref{fig:fc_3GHz} illustrates the generalization performance of different CSI prediction models when evaluated on a test dataset generated at a carrier frequency of \(f_\mathrm{c}=3~\mathrm{GHz}\), while all models were trained exclusively on data generated at \(f_\mathrm{c}=28~\mathrm{GHz}\). This setup induces a noticeable distribution shift in the channel statistics, since lower carrier frequencies exhibit different propagation characteristics and scattering behaviors compared to mmWave frequencies.
The results show that the proposed diffusion-based models achieve the lowest generalization error and maintain robust performance under unseen channel statistics. This advantage stems from modeling the full conditional distribution \(p(\mathbf{H}_\mathrm{f}\mid \mathbf{H}_\mathrm{p})\) rather than learning a single deterministic mapping from past to future CSI. By capturing the underlying spatiotemporal structure of the channel—rather than memorizing training-specific patterns—diffusion-based predictors remain resilient under substantial domain shifts.

The baseline discriminative models suffer from noticeable performance degradation under domain shift. LinFormer in particular exhibits a sharp NMSE increase, indicating strong overfitting to the training distribution at \(f_\mathrm{c}=28~\mathrm{GHz}\) when evaluated at \(f_\mathrm{c}=3~\mathrm{GHz}\). The 3D U-Net diffusion model also exhibits some performance loss due to its relatively high parameter count, which makes it more susceptible to overfitting. By comparison, the ConvLSTM model generalizes better than LinFormer, largely owing to its simpler architecture and lower capacity, which reduce its sensitivity to domain shifts. Overall, these results highlight that diffusion-based CSI predictors—particularly those with lightweight backbone architectures—are substantially more robust to domain shifts, making them especially promising for deployment in real-world wireless systems.

The right plot presents a qualitative comparison of predicted CSI sequences at $\mathrm{SNR} = 10~\mathrm{dB}$ across six forecasting models: DiU, ConvLSTM, GRU, DiT, LinFormer, and DiU3. The first row shows the ground-truth CSI frames, followed by the noisy input frames in the second row. The subsequent rows present the reconstructed CSI sequences for each prediction model, where each column corresponds to a future prediction step. As observed, LinFormer struggles to recover the spatial structure of the wireless channel due to its poor generalization capability, while the other models are able to reconstruct CSI samples more faithfully across prediction steps. Furthermore, DiU3 produces slightly blurred reconstructions, reflecting its performance loss when evaluated under different channel statistics.

\begin{table}[t]
\centering
\caption{Model complexity and approximate inference cost.}
\label{tab:model_complexity}
\small
\setlength{\tabcolsep}{3.pt}
\renewcommand{\arraystretch}{1.15}

\begin{tabularx}{\columnwidth}{@{}l *{3}{>{\centering\arraybackslash}X} @{}}
\toprule
\textbf{Model} & \textbf{FLOPs (G)} & \textbf{Parameters (M)} & \textbf{Time (ms)} \\
\midrule
DiU         & 10.54 & 1.60  & 3.93   \\
DiU-seq2seq & 0.49  & 1.03  & 0.33   \\
DiT         & 0.79  & 0.16  & 4.51   \\
DiU3        & 48.32 & 10.19 & 16.19  \\
LinFusion   & 1.01  & 4.60  & 0.48   \\
LinFormer   & 0.11  & 3.70  & 0.03   \\
ConvLSTM    & 1.52  & 0.15  & 0.24   \\
GRU         & 0.31  & 2.10  & 0.06   \\
\bottomrule
\end{tabularx}
\end{table}

Table~\ref{tab:model_complexity} summarizes the computational complexity of the evaluated CSI prediction models. Specifically, we report the number of floating-point operations (FLOPs) per forward pass, the total number of learnable parameters, and the average wall-clock inference time for a representative setting with \(N_\mathrm{p}=30\) input frames and \(N_\mathrm{f}=10\) predicted CSI frames. 
The wall-clock inference time is measured on an NVIDIA Tesla V100 GPU under a unified implementation and software environment.

Among the baseline methods, GRU and LinFormer exhibit the lowest computational cost. However, both models require a relatively large number of parameters to maintain sufficient representational capacity, which may increase the risk of overfitting. Furthermore, in the reported wall-clock inference time for LinFormer, we assume that the output length is fixed to match the trained model. As a result, predictions are generated in a single forward pass without requiring AR decoding, which significantly reduces the reported inference latency.

AR models generally incur higher computational cost due to temporal recurrence, but offer greater flexibility in handling varying context lengths and prediction horizons. Diffusion-based seq2seq models, particularly DiU-seq2seq, achieve lower inference latency and reduced per-pass computation among the diffusion-based approaches. On the other hand, the diffusion-based predictor with a 3D U-Net backbone incurs substantially higher FLOPs, as it performs joint spatio-temporal processing over the full 3D CSI volume.

\section{Conclusion}\label{sec5}
In this paper, we presented a class of diffusion-based models for CSI prediction, designed to flexibly integrate different time-series forecasting architectures. By progressively learning the underlying distribution of spatial CSI patterns and their temporal evolution, diffusion models capture the inherent uncertainty of wireless channels rather than memorizing a fixed input–output mapping. We considered both designs with and without an explicit temporal encoder: in the former, the temporal encoder extracts latent temporal dynamics from historical CSI, while in the latter, the diffusion generator directly leverages historical CSI samples to predict future frames. Furthermore, we investigated AR and seq2seq inference schemes, explored multiple generator backbones including U-Net, DiT, and 3D U-Net, and incorporated advanced techniques such as DDIM for efficient sampling.
Extensive simulations with the CDL channel model in a mmWave MIMO setup demonstrate that the proposed diffusion-based models consistently outperform state-of-the-art benchmarks such as GRU, ConvLSTM, and LinFormer across diverse SNR levels, prediction horizons, and user mobility scenarios. In particular, the diffusion model with a U-Net backbone delivers the best overall performance, achieving up to $5$--$8$~dB NMSE gains for short-term prediction horizons at high SNRs. Moreover, we show that as few as three diffusion sampling steps are sufficient to attain competitive prediction accuracy, and that generative diffusion models exhibit stronger generalization capabilities than discriminative baselines.

Incorporating explicit physics-inspired modeling into CSI prediction is a promising research direction. While the proposed diffusion-based predictor implicitly captures physical channel characteristics through data-driven learning from standardized channel models, integrating physics-aware priors or parametric channel structures could further enhance prediction accuracy and ensure stronger physical interpretability. In addition, although the DDIM sampler adopted in this work significantly reduces the inference complexity compared with conventional DDPM sampling, other efficient generative frameworks such as flow matching and related few-step generative methods may provide further reductions in sampling latency. Exploring such approaches for low-complexity and real-time CSI prediction constitutes an interesting direction for future research.



\appendices
\section{ConvLSTM}\label{appendix: ConvLSTM}
ConvLSTM extends the standard LSTM architecture by replacing fully connected operations with 2D convolutional operations, enabling it to better preserve local spatial correlations in structured data such as images, videos, or CSI. By sliding convolutional filters across spatial dimensions, ConvLSTM is able to jointly capture spatial features and temporal dependencies more effectively than traditional LSTMs.

Similar to the standard LSTM, ConvLSTM consists of three gates: the input gate, forget gate, and output gate. However, unlike fully connected LSTMs, the computations are performed using convolutions instead of matrix multiplications. The cell state and hidden state at time step \(n\) are updated as follows:
\begin{align}
    \mathbf{Y}_n &= \left[ \mathbf{X}_n, \mathbf{Z}_{n-1} \right], \\
    \left[ \mathbf{i}_n, \mathbf{f}_n, \mathbf{o}_n, \mathbf{g}_n \right] &= \text{Conv2D}(\mathbf{Y}_n), \\
    \mathbf{i}_n &= \sigma(\mathbf{i}_n), \quad
    \mathbf{f}_n = \sigma(\mathbf{f}_n), \quad
    \mathbf{o}_n = \sigma(\mathbf{o}_n), \\
    \mathbf{g}_n &= \tanh(\mathbf{g}_n), \\
    \mathbf{S}_n &= \mathbf{f}_n \odot \mathbf{S}_{n-1} + \mathbf{i}_n \odot \mathbf{g}_n, \\
    \mathbf{Z}_n &= \mathbf{o}_n \odot \tanh(\mathbf{S}_n),
\end{align}
where \( \sigma(\cdot) \) denotes the sigmoid activation function, \( \tanh(\cdot) \) is the hyperbolic tangent function, and \( \odot \) represents element-wise multiplication. The operator \( \text{Conv2D}(\cdot) \) applies a 2D convolution over the spatial dimensions. Here, \(\mathbf{X}_n\) is the input at time step \(n\), \(\mathbf{Z}_{n-1}\) is the previous hidden state, and \(\mathbf{S}_n\) denotes the current cell state.

\section{U-Net}\label{appendix: Unet}

U-Nets were originally introduced for biomedical image segmentation in \cite{ronneberger2015unetconvolutionalnetworksbiomedical}
and have since become a standard denoising backbone in diffusion-based generative models due to their strong ability to capture hierarchical spatial features. The U-Net architecture follows an encoder–decoder design, where the encoder progressively downsamples the input to extract latent representations, while the decoder upsamples and reconstructs the output using both learned features and skip connections from the encoder. 

The network begins with an initial $3\times3$ convolution that projects the two input channels
(real and imaginary CSI components) into $32$ feature channels. These features then enter the encoder path.
To incorporate temporal conditioning on the diffusion step $t$, we embed the scalar timestep using sinusoidal
positional encodings, followed by a two-layer multilayer perceptron (MLP) that maps the embedding to a
$256$-dimensional vector. This timestep embedding is injected into every residual block via adaptive
conditioning, allowing the denoising process to depend on the current stage of the diffusion process.

The encoder consists of two resolution stages.  
The first down block is a standard convolutional block operating on $32$ channels. It applies two residual
blocks, each conditioned on the timestep embedding, followed by a $2\times2$ spatial downsampling
that reduces the resolution from $N_\mathrm{t} \times N_\mathrm{c}$ to $N_\mathrm{t}/2 \times N_\mathrm{c}/2$.  
The second down block is an attention-augmented convolutional block that increases the channel width to
$64$. It applies two residual blocks, each followed by a self-attention layer with a single-head attention mechanism. 
The outputs of these down blocks are stored and later used as skip connections to improve reconstruction fidelity.

At the bottleneck, the network operates on $64$ channels at a spatial resolution of $N_\mathrm{t}/2 \times N_\mathrm{c}/2$.
This mid block consists of a residual block, followed by a self-attention layer, and another residual block.
These layers are also conditioned on the timestep embedding. Including attention at the bottleneck enables
the model to capture global dependencies between antennas and subcarriers.

The first up block is an attention-augmented up block that merges bottleneck features with the skip connection
from the second encoder stage, applying three sequences of residual and attention layers.
The second up block is a standard up block that fuses features with the first encoder skip connection,
applies three residual blocks without attention, and upsamples back to the original $N_\mathrm{t} \times N_\mathrm{c}$ spatial resolution.

Throughout the network, the residual blocks use the Sigmoid Linear Unit (SiLU) activation function.
Specifically, every residual block inside the encoder, bottleneck, and decoder applies the sequence:
Group Normalization $\rightarrow$ SiLU activation $\rightarrow$ Convolution.
Additionally, after the final decoder stage, the features pass through a group normalization layer and
another SiLU activation before the concluding $3\times3$ convolution, where no activation function is applied after the final convolution.

\section{DiT}\label{appendix: DiT}

DiT adopts a Vision Transformer (ViT)-style architecture and incorporates timestep conditioning via adaptive LayerNorm-Zero (adaLN-Zero). The overall architecture includes patch embedding, positional embedding, timestep embedding, and a Transformer block with specialized adaptive layer norm modulation. In particular DiT is similar to a ViT design that processes 2D images as a sequence of patch embedding with learnable positional encoding and stacking Transformer encoder blocks. DiT further applies embedding to diffusion timesteps using sinusoidal functions and an MLP. It also considers the conditioning input to every Transformer block using adaLN-Zero.

 Let the noisy input at diffusion step \(t\) be denoted as \(\mathbf{\tilde{H}}_t \in \mathbb{R}^{ 2 \times N_\mathrm{t} \times N_\mathrm{c}}\). The model operates on non-overlapping patches of size \(P \times P\) with \(P=4\). A strided convolution with kernel and stride equal to \(P\) performs the patchification and per-patch projection, yielding a sequence of tokens \(\mathbf{Z}_0 \in \mathbb{R}^{T \times D}\), where \(T = (N_\mathrm{t}/P)\cdot(N_\mathrm{c}/P)\) is the number of patches and \(D=128\) is the hidden size. To preserve spatial arrangement after flattening, a learnable positional embedding \(\mathbf{P} \in \mathbb{R}^{1 \times T \times D}\) is added to the patch embeddings, giving \(\mathbf{X}_0 = \mathbf{Z}_0 + \mathbf{P}\).

The scalar diffusion index \(t\) is mapped into a high-dimensional representation using sinusoidal functions \(\boldsymbol{\phi}(t) \in \mathbb{R}^{F}\) with \(F=256\). This is subsequently projected through a two-layer MLP with SiLU activation to the model width, producing the timestep embedding \(\mathbf{c} \in \mathbb{R}^{B \times D}\). This embedding conditions every block through adaLN-Zero.

The Transformer stack is composed of \(L=8\) DiT blocks. Each block first processes the conditioning vector \(\mathbf{c}\) using a modulation with a single-layer MLP and SiLU activation to produce six sets of vectors: scale, shift, and gate parameters for both the attention and MLP sublayers. Specifically, we obtain \(\left(\boldsymbol{\delta}^{\mathrm{msa}}, \boldsymbol{\gamma}^{\mathrm{msa}}, \mathbf{g}^{\mathrm{msa}}, \boldsymbol{\delta}^{\mathrm{mlp}}, \boldsymbol{\gamma}^{\mathrm{mlp}}, \mathbf{g}^{\mathrm{mlp}}\right)\), all in \(\mathbb{R}^{B \times D}\). Given input tokens \(\mathbf{X}\), the modulation is applied to LayerNorm outputs as
\[
\mathrm{AdaLN}(\mathbf{X};\boldsymbol{\delta},\boldsymbol{\gamma})
= \mathrm{LN}(\mathbf{X}) \odot (1 + \boldsymbol{\gamma}[:,\text{None},:]) + \boldsymbol{\delta}[:,\text{None},:],
\]
where broadcasting is over the token dimension. The block then computes
\begin{align}
\mathbf{U} &= \mathrm{MHSA}\!\left(\mathrm{AdaLN}(\mathbf{X};\boldsymbol{\delta}^{\mathrm{msa}},\boldsymbol{\gamma}^{\mathrm{msa}})\right), \\
\mathbf{X} &\leftarrow \mathbf{X} + \mathbf{g}^{\mathrm{msa}}[:,\text{None},:] \odot \mathbf{U}, \\
\mathbf{V} &= \mathrm{MLP}\!\left(\mathrm{AdaLN}(\mathbf{X};\boldsymbol{\delta}^{\mathrm{mlp}},\boldsymbol{\gamma}^{\mathrm{mlp}})\right), \\
\mathbf{X} &\leftarrow \mathbf{X} + \mathbf{g}^{\mathrm{mlp}}[:,\text{None},:] \odot \mathbf{V}.
\end{align}
The multi-head self-attention employs \(H_a=8\) heads, so each head has width \(D/H_a=16\). The feed-forward MLP expands the hidden dimension to \(\rho D = 256\) with Gaussian Error Linear Units (GELU) nonlinearity before projecting back to \(D\).

After the final block, another adaLN modulation and a linear projection with a single-layer MLP and SiLU activation map each token to \(2P^2 \) output values, corresponding to a patch of size \(P \times P\) with \(C\) channels. An inverse patching step reconstructs the spatial tensor \(\hat{\mathbf{H}} \in \mathbb{R}^{2 \times N_\mathrm{t} \times N_\mathrm{c}}\).

To ensure stable training under high-noise conditions, the weights producing the modulation vectors and residual gates, as well as the final linear projection, are initialized to zero. Consequently, the model behaves as an approximate identity function at initialization. The adopted configuration uses \(P=4\), \(D=128\), \(L=8\), \(H_a=8\), and an MLP expansion ratio of \(\rho=2.0\), which balances global modeling capacity with computational efficiency for CSI prediction tasks.


\section{3D U-Net}\label{appendix:unet3d}

We extend the U-Net architecture to a 3D variant in order to model the spatio-temporal structure of CSI volumes.
The network takes as input a stack of frames consisting of both the observed past and the future frames
concatenated along the temporal dimension, and augments the channel dimension with a binary mask channel
($1$ for past frames and $0$ for future frames) to distinguish between conditioning and prediction targets.

The diffusion timestep $t$ is embedded using sinusoidal position encodings followed by a two-layer MLP that
projects the embedding to a $256$-dimensional vector.
This embedding is injected into every residual block to enable adaptive conditioning on the current stage of the diffusion process.

The encoder begins with an initial 3D convolutional block that projects the $(2{+}1)$ input channels (real, imaginary,
and mask) into $32$ feature channels. It then applies three downsampling stages with output widths of $64$, $128$, and $256$, respectively. Each 3D convolutional block contains
two $3{\times}3{\times}3$ convolutions, each followed by Group Normalization and SiLU activations, with the timestep
embedding added between the two convolutions. After each stage, the spatial resolution is reduced by half via maxpooling, while preserving the temporal dimension.
Skip connections are extracted from the outputs of each 3D convolutional block before pooling for use in the decoder.

At the bottleneck, the network applies another 3D convolutional block with $256$ channels, followed by a self-attention layer that performs multi-head self-attention only along the temporal axis while
treating each spatial location independently. This allows the model to explicitly capture temporal dependencies between CSI frames while keeping the computational cost independent of the spatial size.

The decoder mirrors the encoder and consists of three upsampling stages.
Each block begins with a 3D transposed convolutional block layer that upsamples only the spatial dimensions, concatenates
the result with the corresponding skip connection, and processes the combined features to fuse spatial and temporal information. The channel dimensions are reduced progressively
from $256 \rightarrow 128 \rightarrow 64 \rightarrow 32$, while the temporal resolution remains constant.

Finally, a $1{\times}1{\times}1$ convolution maps the $32$ feature channels to the two output channels
representing the real and imaginary parts of the predicted future CSI. Since the input volume includes both
past and future frames, only the last frames of the output are retained to produce predictions.

Within every 3D convolutional block, the operations are applied in the following order:
Convolution $\rightarrow$ Group Normalization $\rightarrow$ time-embedding addition $\rightarrow$ SiLU $\rightarrow$
Convolution $\rightarrow$ Group Normalization $\rightarrow$ SiLU. Thus, the SiLU activation is consistently used in
all residual blocks across the encoder, bottleneck, and decoder. The multi-head self-attention layer does not
use any activation internally and relies on the surrounding residual blocks for nonlinearity. No activation is
applied after the final convolution to allow the network to produce unconstrained outputs.

\bibliographystyle{IEEEtran}
\bibliography{Refs}

\begin{IEEEbiography}[{\includegraphics[width=1in, height=1.25in,trim=5cm 0cm 15cm 0cm, clip]{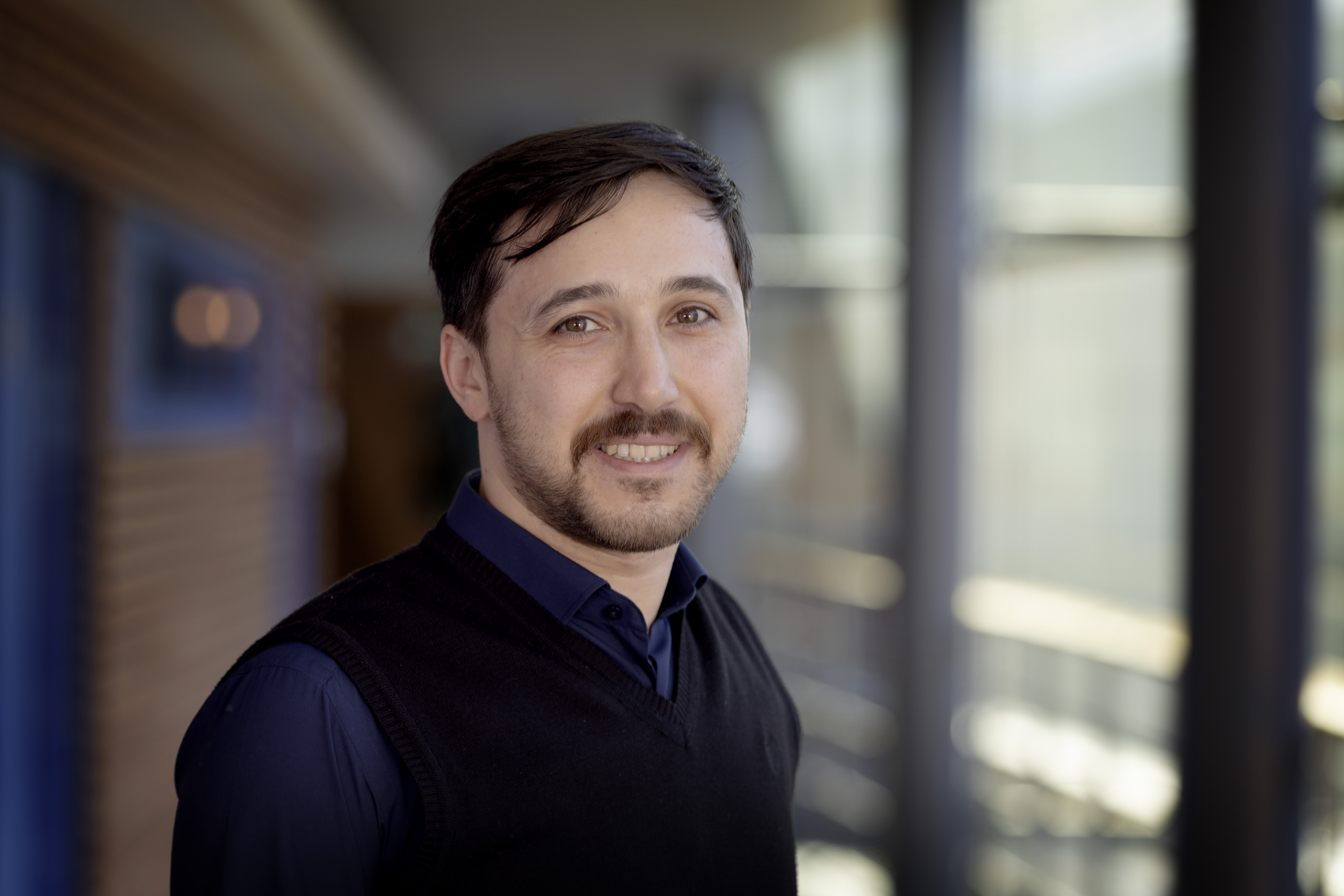}}]{Mehdi Sattari}
received the B.Sc. degree from the Iran University of Science and Technology and the M.Sc. degree from the University of Tehran. He received the Ph.D. degree in Electrical Engineering from Chalmers University of Technology, Gothenburg, Sweden. He was a Visiting Researcher with Imperial College London, London, U.K., in 2023. He is currently a Postdoctoral Researcher with the Department of Electrical Engineering, Chalmers University of Technology. His research interests are deep learning and generative AI for wireless communications, particularly CSI estimation, compression, and prediction.
\end{IEEEbiography}

\begin{IEEEbiography}[{\includegraphics[width=1in,height=1.25in,clip,keepaspectratio]{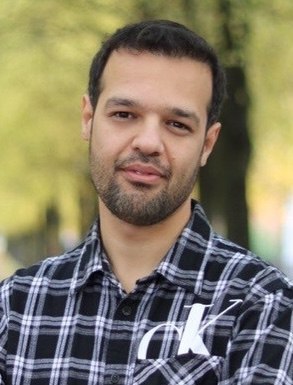}}]
{Javad Aliakbari}
received the B.Sc. degree in electrical engineering from the University of Tehran and the M.Sc. degree in communication systems from Sharif University of Technology, Tehran, Iran. He is currently pursuing the Ph.D. degree at Chalmers University of Technology, Gothenburg, Sweden. Before starting his Ph.D., he worked in industry as a data scientist and researcher in machine learning, signal processing, and natural language processing. His research interests include graph neural networks, federated learning, privacy-preserving machine learning, and generative models.
\end{IEEEbiography}

\begin{IEEEbiography}[{\includegraphics[width=1in,height=1.45in,clip,keepaspectratio]{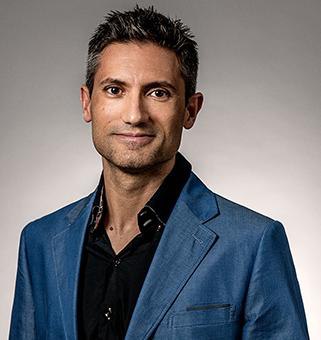}}]
{Alexandre Graell i Amat}
received the M.Sc. degree in electrical engineering from the Politecnico di Torino, Turin, Italy, in 2000, the M.Sc. degree in telecommunications engineering from the Universitat Politècnica de Catalunya, Barcelona, Catalonia, Spain, in 2001, and the Ph.D. degree in electrical engineering from the Politecnico di Torino, in 2004. From 2001 to 2002, he was a Visiting Scholar with the University of California at San Diego, San Diego, CA, USA. From 2001 to 2004, he held a part-time appointment with the STMicroelectronics Data Storage Division, Milan, Italy, as a Consultant on coding for magnetic recording channels. From 2002 to 2003, he held a visiting appointment with Universitat Pompeu Fabra, Barcelona, and the Telecommunications Technological Center of Catalonia, Barcelona. From 2004 to 2005, he was a Visiting Professor with Universitat Pompeu Fabra. From 2006 to 2010, he was with the Department of Electronics, IMT Atlantique (formerly ENST Bretagne), Brest, France. Since 2019, he has been an Adjunct Research Scientist with Simula UiB, Bergen, Norway. He is currently a Professor with the Department of Electrical Engineering, Chalmers University of Technology, Gothenburg, Sweden. His research interests include coding theory with application to distributed learning and computing, storage, privacy and security, and communications. He received the Marie Skłodowska-Curie Fellowship from European Commission and the Juan de la Cierva Fellowship from Spanish Ministry of Education and Science. He received the IEEE Communications Society 2010 Europe, Middle East, and Africa Region Outstanding Young Researcher Award. He was the General Co-Chair of the 7th International Symposium on Turbo Codes and Iterative Information Processing, Sweden, in 2012, and the TPC Co-Chair of the 11th International Symposium on Topics in Coding, Canada, in 2021. He was an Associate Editor of IEEE Communications Letters from 2011 to 2013. He was an Associate Editor and the Editor-at-Large of IEEE Transactions on Communications from 2011 to 2016 and from 2017 to 2020. He is an Area Editor of IEEE Transactions on Communications.
\end{IEEEbiography}

\begin{IEEEbiography}[{\includegraphics[width=1in,height=1.25in,clip,keepaspectratio]{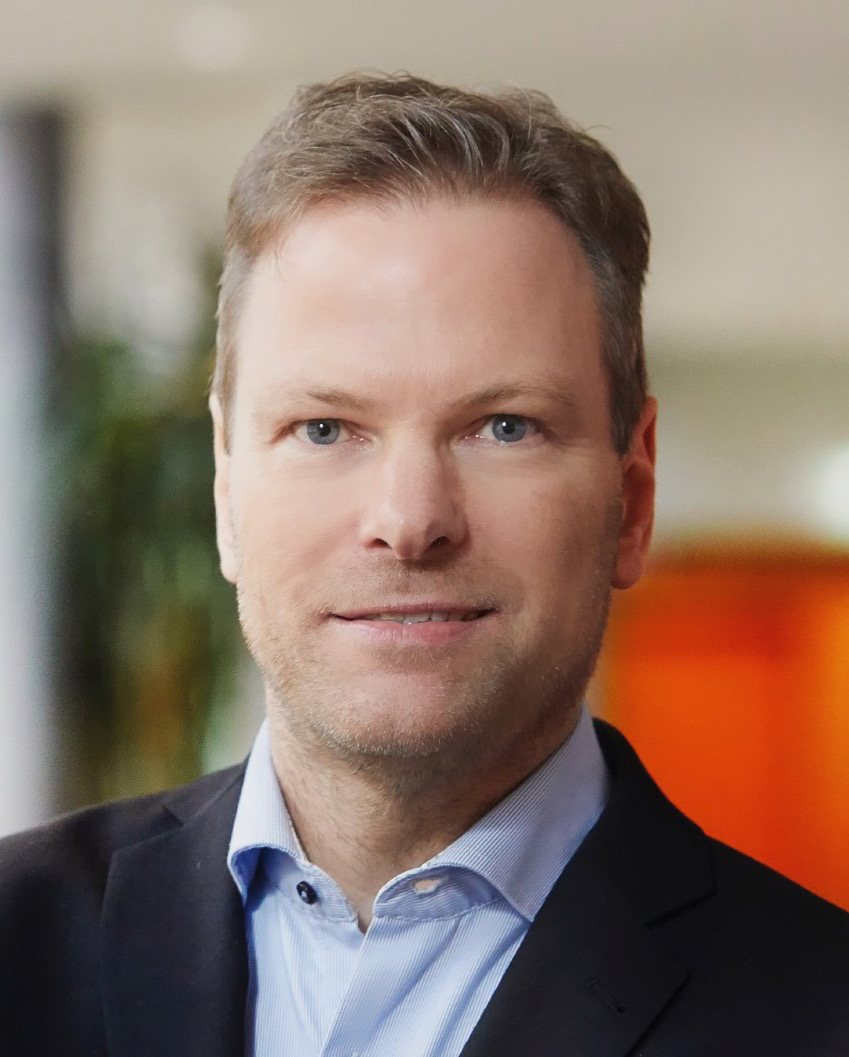}}]
{Tommy Svensson}
 is Full Professor in Communication Systems at Chalmers University of Technology in Gothenburg, Sweden, where he is leading the Wireless Systems research on air interface and wireless backhaul networking technologies for future wireless systems. He received a Ph.D. in Information theory from Chalmers in 2003, and he has worked at Ericsson AB with core networks, radio access networks, and microwave transmission products.

He was involved in the European WINNER I/II/+ and ARTIST4G projects that made important contributions to the 3GPP LTE standards, the EU FP7 METIS and the EU H2020 5GPPP mmMAGIC and 5GCar projects towards 5G, and the Hexa-X-I/II, RISE-6G, SEMANTIC, ROBUST-6G and ECO-eNET projects towards 6G, as well as in the ChaseOn/Bridge Center/emerging WiTECH antenna systems excellence centers at Chalmers targeting mm-wave and (sub)-THz solutions for 5G/6G access, backhaul/ fronthaul and V2X scenarios.

His main research interests are in design and analysis of mobile communication systems, physical layer algorithms, multiple access, resource allocation, cooperative/ situational-aware communications, mm-wave/ sub-THz communications, C-V2X, ISAC, physical-layer security, non-terrestrial-networks, sustainable design, end-to-end architecture.

He has co-authored 7 books, 150 journal papers, 180 conference papers, and 80 public EU projects deliverables. He is founding editorial board member and editor of IEEE JSAC Series on Machine Learning in Communications and Networks, has been Chairman of the awards winning IEEE Sweden joint Vehicular Technology/ Communications/ Information Theory Societies chapter, editor of IEEE Transactions on Wireless Communications, IEEE Wireless Communications Letters, Guest editor of several top journals, organized several tutorials and workshops at top IEEE conferences, Lead local organizer of EuCNC \& 6G Summit 2023, and served as coordinator of the Communication Engineering Master’s Program at Chalmers. He is leading the Swedish VR Research Environment “Foundational Algorithms, Protocols, and Systems for Multi-Tier 6G-Non-Terrestrial Networks Integrated Communication and Environmental Sensing (6G-NTN-E)” that started in 2025, and since 2022 he is board member of Swedish telecommunications regulator (PTS).
\end{IEEEbiography}

\end{document}